\documentclass[twocolumn]{aastex631}

\usepackage{enumitem}
\setlist{nosep}

\usepackage{amsmath}
\usepackage{array}
\usepackage{soul}
\shorttitle{Cluster Membership and Variability of Mon R2 YSOs}
\shortauthors{Jiang \& Hillenbrand, 2023}

\begin{document}

\title{The Emerging Stellar Complex in Mon R2: Membership and Optical Variability Classification}

\author[0000-0003-1369-602X]{Sally D. Jiang}
\affiliation{Department of Astronomy, Yale University, New Haven, CT 06511, USA}
\affiliation{Department of Astronomy, Columbia University, New York, NY 10027, USA}

\author{Lynne A. Hillenbrand}
\affiliation{Department of Astronomy, California Institute of Technology, Pasadena, CA 91125, USA}

\begin{abstract}
Monoceros R2 (Mon R2) is one of the closest large active star-forming regions. 
This extremely young and partially embedded region
provides an excellent laboratory for studying star formation and the early evolution of young stellar objects (YSOs). 
In this paper, we conduct an optical study of the greater Mon R2 region.
Beginning with 1690 previously identified candidate YSOs,
we used 496 sources with good proper motions and parallaxes from Gaia DR3 to determine the astrometric properties for likely members of Mon R2.
We then used both astrometric and photometric (isochronal and variability) criteria to determine that 308 of these stars are highly probable members. 
Using the same criteria, we considered a broad area search around Mon R2 in Gaia DR3, and separated candidate members from field stars.
In total, we selected 651 likely new cluster members that had been missed in the previous x-ray and infrared excess selection techniques used in the past to establish cluster membership. Revised astrometric properties of the cluster were found using the combined sample of $\sim959$ highly probable member stars.  For the literature plus new candidate member list, optical light curves were compiled from the Zwicky Transient Facility. 
For 470 
identified variable sources, we attempted classification based on the Flux Asymmetry (M) and Quasi-Periodicity (Q) metrics. We find that Mon R2 is dominated by quasi-periodic symmetric variables, with aperiodic sources also a significant population.  A few tens of large-amplitude variables are identified that may be of interest for further study. 

\end{abstract}


\section{Introduction} \label{sec:intro}

Stars form from the collapse and fragmentation of molecular clouds. These dust-rich gas clouds have enough material to produce anywhere from tens to tens of thousands of stars, often resulting in dense stellar systems known as star clusters.
Detailed study of young stellar populations, and their clustering properties, in such active regions     
can help inform the processes and conditions during star formation from dense interstellar material. 
Before nuclear fusion begins within its core, a forming star goes through core collapse, a main accretion phase, and finally slow contraction, eventually reaching a quasi-equilibrium. During the early stages of a star’s pre-main sequence evolution, the object is denoted as a Young Stellar Object (YSO). 

One defining property of YSOs is their spectroscopic and photometric variability \citep{joy1945}. 
As a protostar with a circumstellar disk, the object experiences many dynamic phenomena including accretion of material, ejection of material in outflows and jets, magnetic activity manifest as cool and hot spots on the surface, and occasional outbursts \citep{semkov2011,fischer2023}. These physical phenomena each result in differing amplitudes and timescales of the star’s variability and cause brightness fluctuations in different regions in the electromagnetic spectrum. Here, we study the optical variability of YSOs toward the Mon R2 region.

The Monoceros R2 Molecular Cloud, 
often referred to as just ``Mon R2", is one of the closest large active star-forming regions from the sun. The traditional distance estimate of $830\pm50$ pc \citep{racine1968,herbstracine1976} was updated based on high-precision VLBA measurements to $893\pm42$ pc by \cite{dzib2016}. The Monoceros R2 region contains several early-type stars and reflection nebulae, and deeply embedded star clusters \citep[see][for a review]{carpenter2008}. It also contains the YSO outburst source V899 Mon, an EX Lup type object.

The main star cluster in Mon R2 was first identified by \cite{beckwith1976} and further revealed by \cite{hodapp1994}. Although embedded, it is associated with NGC 2170. A second dense embedded cluster associated with GGD 12-15 is just to the east \citep{gutermuth2005}, and a third associated with GGD 16-17 is even further east\footnote{The GGD numbers are Herbig-Haro objects from the catalog of \cite{ggd1978}.}. A network of filamentary gas and dust structures spiral outwards from the central cluster \citep{rayner2017,trevino2019,kumar2022}. In this and other examples of hub-and-spoke type structure in the dense interstellar medium, it is thought that the hub maintains star formation through the accretion of material that flows from the filaments to the central hub \citep{rayner2017}, and that these central regions are more likely to produce massive stars and clusters, relative to star formation in the fragmenting filaments. Other examples of hub-filament systems are presented in e.g. \cite{mookerjea2023}, \cite{Nagai1998}, \cite{Myers2009}, \cite{Wang2020ApJ...905..158W}.

Whereas most of the previous literature on Mon R2 region has presented data and analysis on the gas \citep[e.g.][]{loren1977,pilleri2012,ginard2012,trevino2016,keown2019} or the dust \citep[e.g.][]{pokhrel2016,sokol2019,hwang2022}, 
our focus is on the young stellar population.
The Mon R2 region has a very low inclination on the sky, allowing us to see the entire cluster face-on \citep{kumar2022}. Young embedded star clusters provide an opportunity to study both star formation and early-stage stellar evolution. Using photometric measurements of the large sample of young, bright, and active stars in these clusters, we can better understand the surrounding environment and conditions for star formation. 

Past work on the properties of stars includes that of \citep{carpenter1997,andersen2006} in the main Mon R2 cluster and \citep{maaskant2011} in the GGD 12-15 cluster. However, little examination has been made of stellar properties of the YSOs in the larger region beyond the main clusters. Detailed work beyond census building \citep[e.g.][]{gutermuth2009}, specifically the establishment of membership of candidate young stars, has not been done. Nor has there been a large-scale analysis of optical variability using long-term light curves.
Mid-infrared variability in the main cluster was characterized by \cite{rebull2014},
and a recent preprint by \cite{Orcajo2023} studies optical variability on relatively short timescales. 

In this paper, we reconsider the census of the greater Mon R2 region, starting with a star list based on the existing literature. We then use Gaia DR3 astrometry and photometry \citep{gaia2022}, and light curves from the Zwicky Transient Facility \citep[ZTF; ][]{2019PASP..131f8003B},  to firmly establish the Mon R2 census.
We then study the optical photometric variability using the ZTF light curves, which sample timescales from days to years, allowing us to capture the short-term and long-term variability of the YSOs. Using thresholding metrics, we then test whether each star’s brightness over time is changing and hence whether it is variable or not. After identifying variable stars in Mon R2 we can then classify the variability properties based on established variability metrics, which can suggest the physical phenomena that are causing the lightcurves to vary. Section \ref{sec:data} describes the assembled data sets and our data culling processes and quality control steps. Section \ref{sec:membership} details how we assess cluster membership. Section \ref{sec:opticalvar} presents our analysis of the optical light curves. Section \ref{sec:discussion} contains our discussion and conclusions.

\section{Data and Data Vetting} \label{sec:data}

\begin{figure*}[ht]
    \centering
    \includegraphics[width=1.1\linewidth]{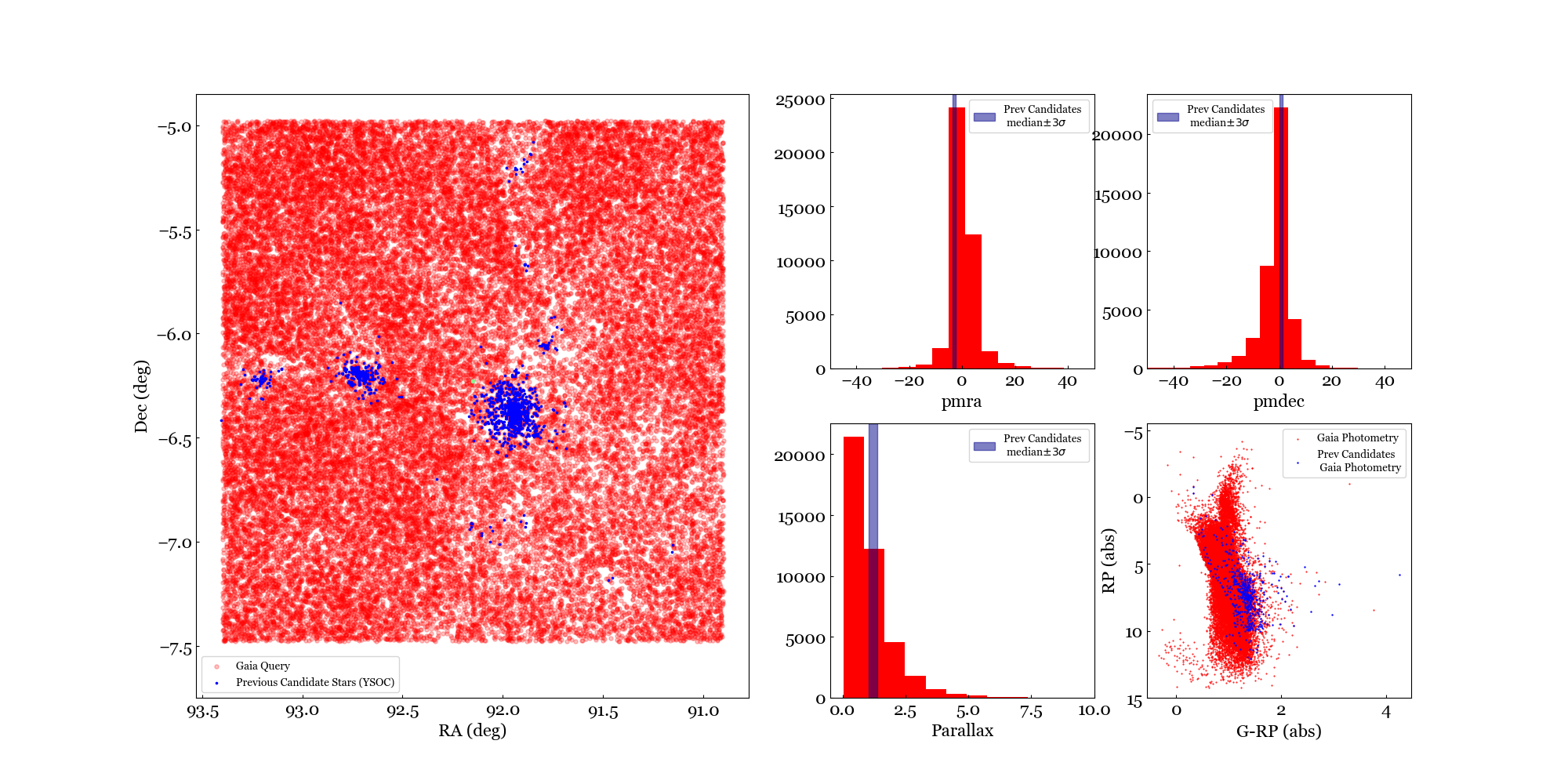}
    \caption{Left: Spatial distribution of sources considered in this study.  Blue points show stars previously associated with the Mon R2 region from infrared and x-ray surveys, that are mainly concentrated in the central cluster and the two groupings to the east: GGD 12-15 and GGD 16-17. 
    Red points result from our initial Gaia query and show the broad distribution of the optical sources that we tested for potential cluster membership. 
    Right: histograms in proper motion RA, proper motion Dec, and parallax with red indicating the Gaia stars and blue bars showing the $\pm 3\sigma$ range of the previously associated stars. Lower right panel is a color magnitude diagram in Gaia G-RP versus RP filters for the same two populations.    }
    \label{fig:initialmem}    
\end{figure*}

\subsection{YSOC: Young Stellar Objects Previously Associated with Mon R2} \label{sec:YSOCDataset}
As described in \cite{hillenbrand2021},
YSOC (the Young Stellar Object Corral) is an online database of nearby young stars containing astrometric, kinematic, photometric, spectroscopic, and other observed stellar properties, as well as associated membership information. The data is compiled from various papers studying young stellar objects from x-rays to millimeter wavelelengths, and complemented with data from survey archives, The site contains stars associated with various different regions, such as the North American-Pelican Nebula, Orion, Ophiuchus, etc. 

For Mon R2, the historical literature that has established the stellar census is mainly based on infrared observations, namely the near-infrared surveys from the ground of \cite{carpenter1997} and from HST of \cite{andersen2006}, and the mid-infrared surveys with Spitzer of \cite{gutermuth2009} and \cite{kryukova2012}. In addition, X-ray surveys by \cite{kohno2002}, \cite{nakajima2003} and \cite{getman2017} also contributed unique candidate members.
Our initial list of plausible members of the Mon R2 region is the unique list of 1690 stars compiled from the above sources. 

\subsection{Gaia DR3: Astrometry and Optical Photometry}
Gaia is a European Space Agency Mission \citep{gaia2016} 
measuring high-precision astrometric, photometric, and spectroscopic data for over 1.5 billion sources. Gaia Data Release 3 (DR3) \citep{gaia2022} occurred in June 2022 and provided updated position, proper motion, and parallax information for magnitudes as faint as $G\approx 21$ mag, as well as photometry.

A region 2.5 deg $\times$ 2.5 deg in size 
(corresponding to about 36.3 x 36.3 pc$^2$, for our adopted distance of 833 parsecs) was queried in Gaia DR3 around the position RA: 06h08m35.5s, Dec: -06d13m42.60s, resulting in a return of 107,557 stars. This provides the material to search for new members of the greater Mon R2 region that were potentially missed by previous work that relied on less diagnostic selection techniques. 

For our Gaia query, we applied parallax quality cuts to eliminate nonphysical parallax (requiring $plx > 0$) 
and bad parallax measurements (requiring $plxSNR > 2$). 
This left 41,493 stars from our original 107,557 sources. Figure \ref{fig:initialmem} shows the region of Mon R2 we studied, with the sources resulting from the initial Gaia (red) and the sources previously associated with Mon R2 as compiled in YSOC (blue). The right-hand side shows various distributions comparing all Gaia stars in the field with the previously associated stars.

 \subsection{Zwicky Transient Facility: Optical Time Series Photometry}

The Zwicky Transient Facility \citep[ZTF; ][]{2019PASP..131f8003B} is a survey  using the Samuel Oschin 48-inch telescope (P48) at Palomar Observatory.  ZTF catalogs over a billion stars in the northern sky and measures photometric time series, primarily in the r-band ($\lambda_{eff} = 6340$ \AA)  and g-band ($\lambda_{eff} = 4722$ \AA).  Due to the red nature of our sources, we consider only $r$-band lightcurves due to their higher signal-to-noise.  We utilize photometry from ZTF Data Release 12, spanning a 4-year period from March 2018 to May 2022. 

We queried ZTF using the Caltech/IPAC InfraRed Science Archive (IRSA). For stars of interest, we used the RA and Dec coordinates from Gaia or, for sources not present in Gaia, from YSOC (which have their provenance in Gaia then 2MASS then WISE). We searched a 1.5 arcsec radius around each star’s coordinate for ZTF counterparts. The search radius was chosen to minimize contamination in the lightcurve from nearby stars. We checked the quality of the matches by computing the difference of the RA and Dec coordinates of the star, and the median RA and Dec for all apparitions of the source in the ZTF light curve.  After calculating the euclidean separation ($\sqrt{\Delta RA^2 + \Delta Dec^2}$), we examined these differences to validate that each lightcurve corresponds to the intended source. We also examined the stars with separation $>1 \arcsec$ using CDSPortal\footnote{http://cdsportal.u-strasbg.fr} to check for contamination by closely projected stars (e.g. cases where the optically visible ZTF source is not the same object as the embedded claimed Mon R2 member).

Separately, we identified cases of different sources in close proximity to one another having the same lightcurve, and assigned the lightcurve to one of the sources if there was no ambiguity, but eliminated objects where no unique attribution could be made.

The retrieved ZTF light curves were considered in our analysis below only if the source had a median magnitude $r<21$, with $>25$ observations in the time series.
This restricts our analysis to the higher-quality light curves.   We also note that ZTF saturates at approximately 13'th mag, so the brightest few sources in our sample do not have lightcurves.

\section{Cluster Membership Methods and Results} \label{sec:membership}

Ultimately, we are interested in studying the optical light curves of Mon R2 members with ZTF.
In order to establish a membership list, we assembled data and applied cuts designed to leave us with high-quality members. In this section, we describe the steps taken to separate contaminating foreground and background field stars from likely young stellar members of the Mon R2 region. In the subsections below we design and apply criteria based on astrometry, photometry, and photometric variability. We then weight the different membership criteria and select a final list of candidate members from the stars assembled from YSOC and the wide-field Gaia query.

\subsection{Astrometric and Photometric Properties of Mon R2 Established from the Literature Sample}\label{subsec:establishmember}

\begin{figure*}[h]
    \centering
    \includegraphics[width=\linewidth]{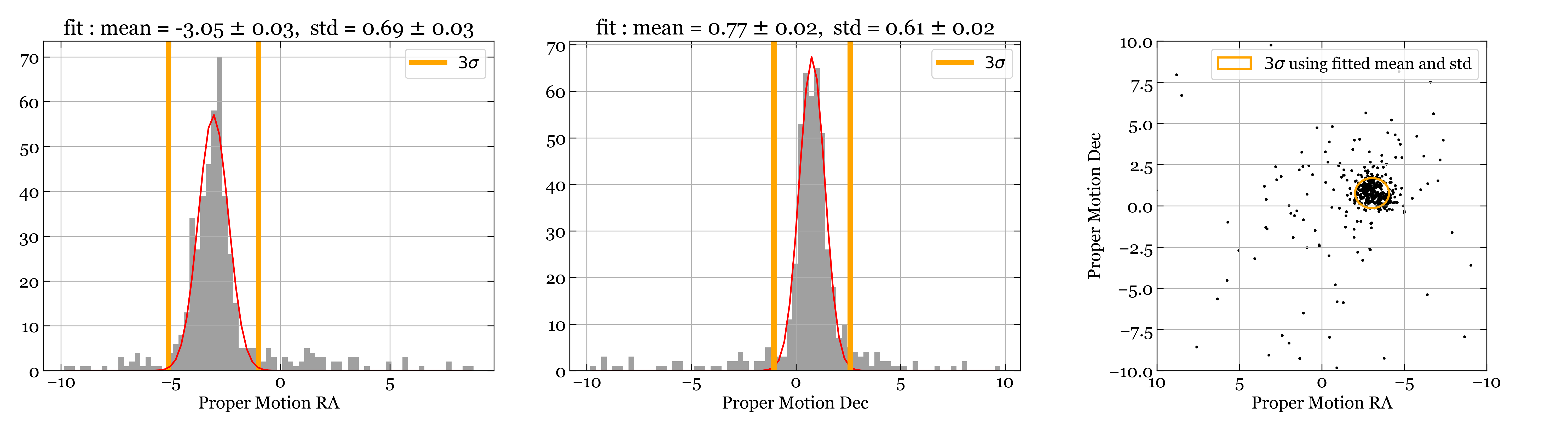}
    \includegraphics[width=\linewidth]{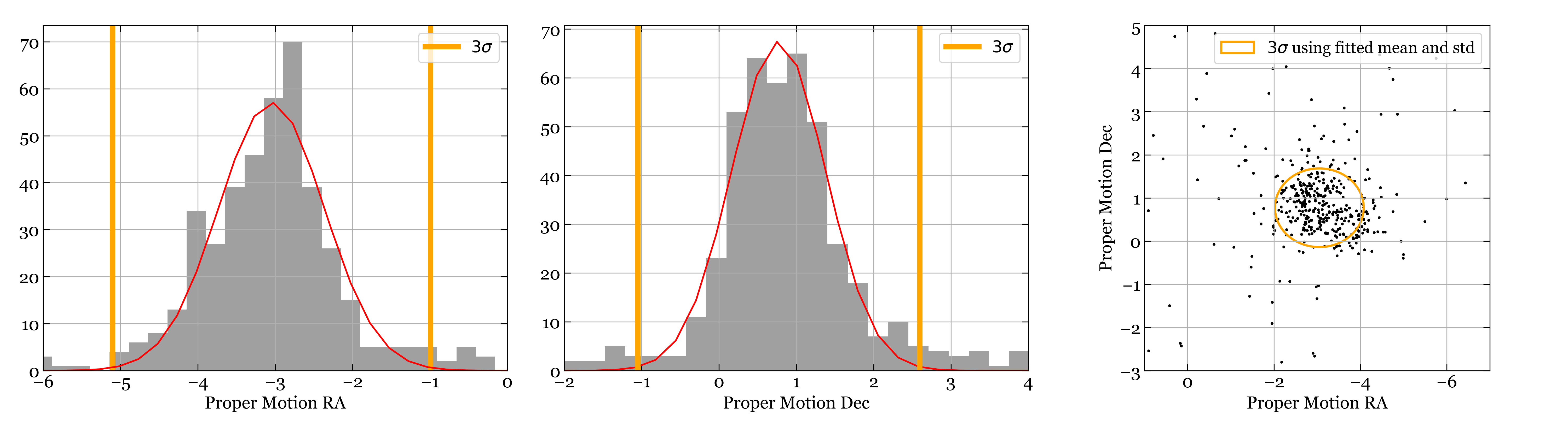}
    \caption{Top: Distribution of proper motion in RA (left) and proper motion in Dec (middle) for YSOC Catalog Stars. Gaussian fits to the distributions are overplotted in red, with $\pm3 \sigma$ values marked with orange lines. The vector point diagram (right) shows the proper motion distribution, with an ellipse indicating the $3\sigma$ contour. Bottom: Same plot as above but zoomed in around the main distribution.}
    \label{fig:pmfit}    
\end{figure*}

\begin{figure}[h]
    \centering
    \includegraphics[width=1.1\linewidth]{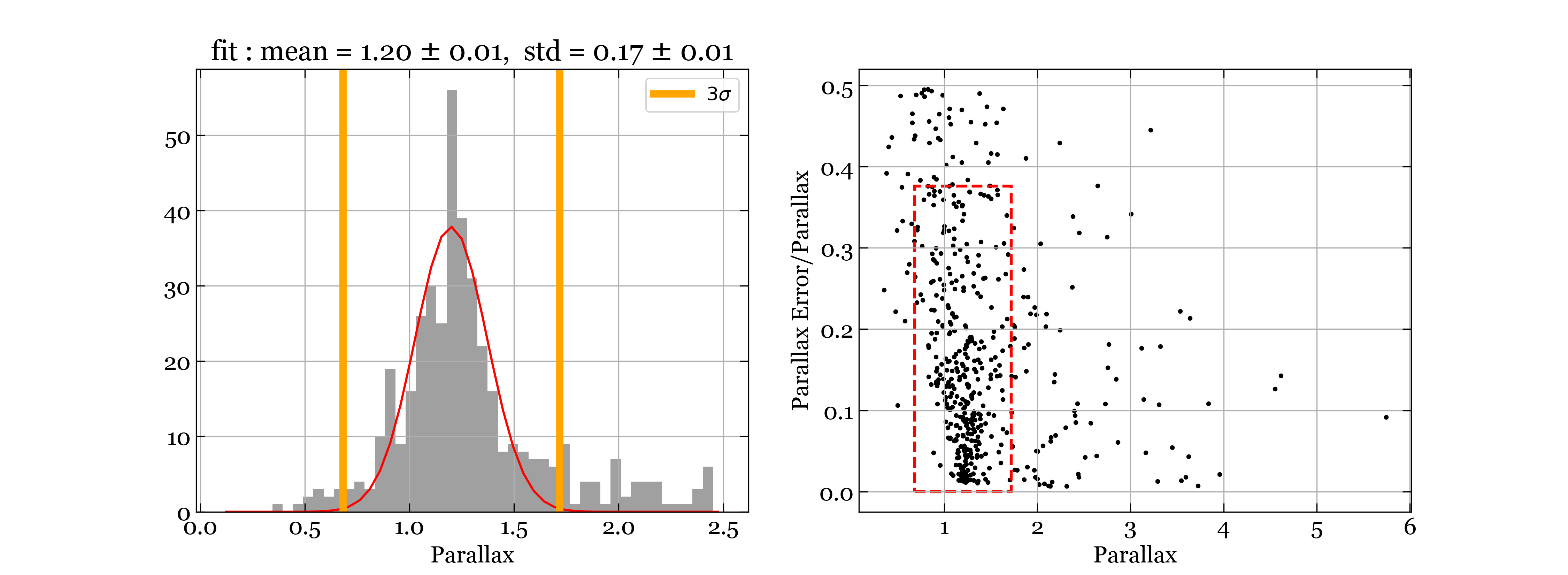}
    \includegraphics[width=1.1\linewidth]{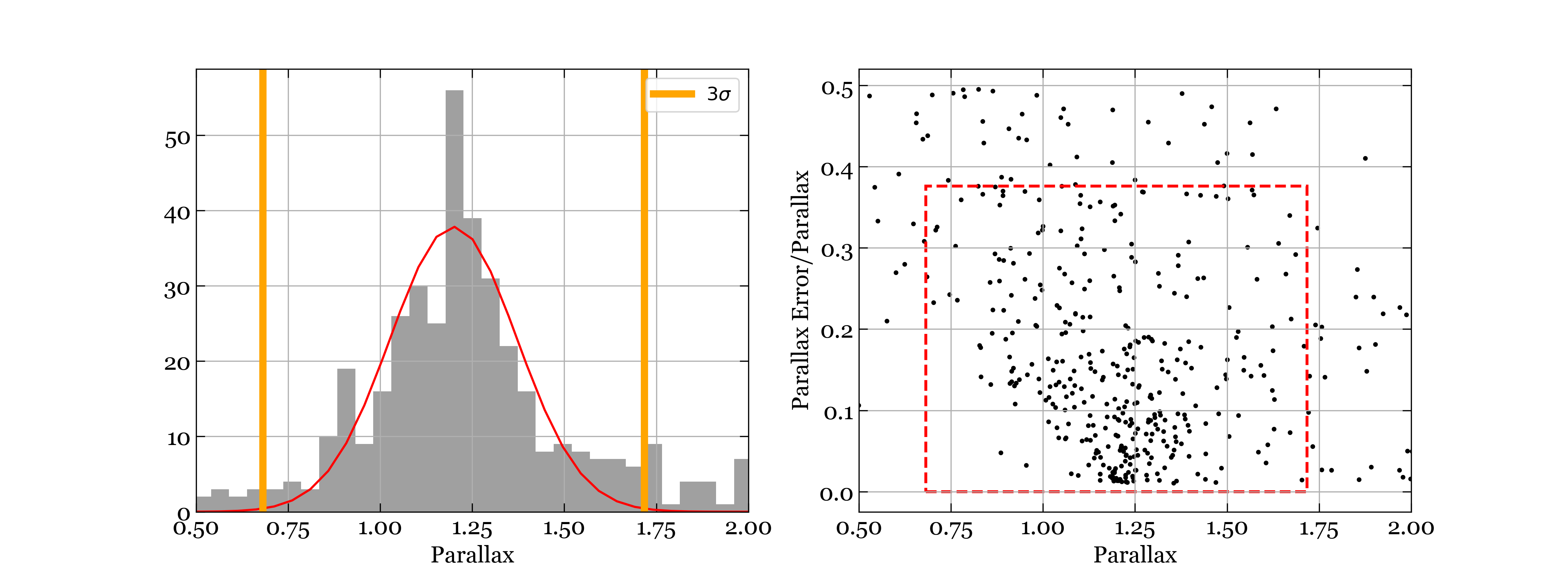}
    \caption{
    Left: Gaussian fit to parallax values of YSOC catalog stars with parallax error over parallax larger than the 90th percentile value (0.375).
    Right: Parallax versus inverse signal-to-noise in the parallax measurements; The boxed region shows data within $3\sigma$ of Gaussian mean. Bottom: Same plot as above but zoomed in around the main distribution.}
    \label{fig:plxfit}    
\end{figure}

 Because star clusters consist of stellar groupings that are spatially and kinematically associated, their members should be located around the same distance from us, and be co-moving in the plane of the sky 
 along similar directions. Using these assumptions, we can assess the parallax and proper motion values of all the stars in the greater Mon R2 field to find stars which are moving coherently, and identify those which have discrepant motions. To establish the astrometric parameters of the cluster we use the list of stars previously associated with Mon R2 as compiled from the YSOC database. 

As stated above, we started with 1690 stars associated in the literature with the Mon R2 Region. We find that 994 of these stars lack proper motions or parallax values in Gaia DR3. This is due to the highly embedded nature of the core (or ``hub") of Mon R2, where most of the members have been identified, as well as high extinction in some of the ``spoke" regions.  The stars without astrometric values in Gaia DR3 are dimmer in optical photometric bands compared to those having high quality astrometry.  Specifically, based on matching to PanSTARRS, which is deeper than ZTF, we find that at least 90\% of the stars with parallax and proper motion measurements have $r< 21$ mag.

We assert that it is reasonable to calculate the cluster parameters using the optically brighter sources in the cluster, but acknowledge potential bias if the more embedded sources have different kinematics. The stars without astrometry are not eliminated from our consideration, as we can still assess their membership through other criteria. 
However, we note that a brightness of $r<21$ mag was also the limit imposed above on the ZTF data set, so the brighter population is in fact the ultimate target of our investigation.

For the remaining 696 stars from YSOC having parallax values in Gaia DR3, we applied parallax quality cuts to eliminate nonphysical parallax (requiring $plx > 0$) and bad parallax measurements (requiring $plxSNR > 2$). This left 496 stars with quality parallax measurements. 
As a first step in creating membership probability criteria to select likely members of MonR2, apart from field stars, we found the mean and modal astrometric values for this sample of likely member stars, 
established from the literature.

To calculate the cluster proper motions, we first plot the one-dimensional distributions of Proper Motion (RA) and Proper Motion (Dec), and fit Gaussian functions to extract the mean and standard deviation (Figure \ref{fig:pmfit}). It should be noted that we fit only data within the range -10 to 10 mas/yr, for both proper motions, which encloses the medians of the proper motions ($\mu_{\alpha, med}$=-2.86, $\mu_{\sigma, med}$=0.70), and avoids stars in the initial catalog with extreme values. From our fits, we find that suggested members of the Mon R2 cluster have mean proper motions and standard deviations:

\begin{itemize}
\centering
    \item{Proper Motion RA: $-3.05 \pm 0.69$ mas/yr}
    \item{Proper Motion Dec: $0.77 \pm 0.61$ mas/yr}
\end{itemize}
with errors of $\pm0.03$ mas/yr in RA and $\pm0.02$ mas/yr in Dec.
For parallax, we use the same Gaussian fitting process
after eliminating stars with parallax error over parallax less than 0.38, or the 90th percentile parallax error over parallax value (Figure \ref{fig:plxfit}b). We thus ensure working with stars with both physical and accurate parallax measurements. Figure \ref{fig:plxfit}a shows the Gaussian fit, which has mean and standard deviation:
\begin{itemize}
\centering
    \item{Parallax: $1.20 \pm 0.17$ mas}
\end{itemize}
with error of $\pm0.01$ mas.
The inferred distance of $833.33 ^{+103.40}_{-137.54}$ parsecs
matches well with previously determined distances to the cluster 
(see references in Sections \ref{sec:intro} and \ref{sec:discussion}).

We emphasize that the above kinematic properties are derived from candidate cluster members identified in previous literature.
In the next subsection, we use the parallaxes and proper motions of individual stars as one factor in assessing their probability of membership in the cluster.
\begin{figure*}[h]
    \centering
    \includegraphics[width=\linewidth]{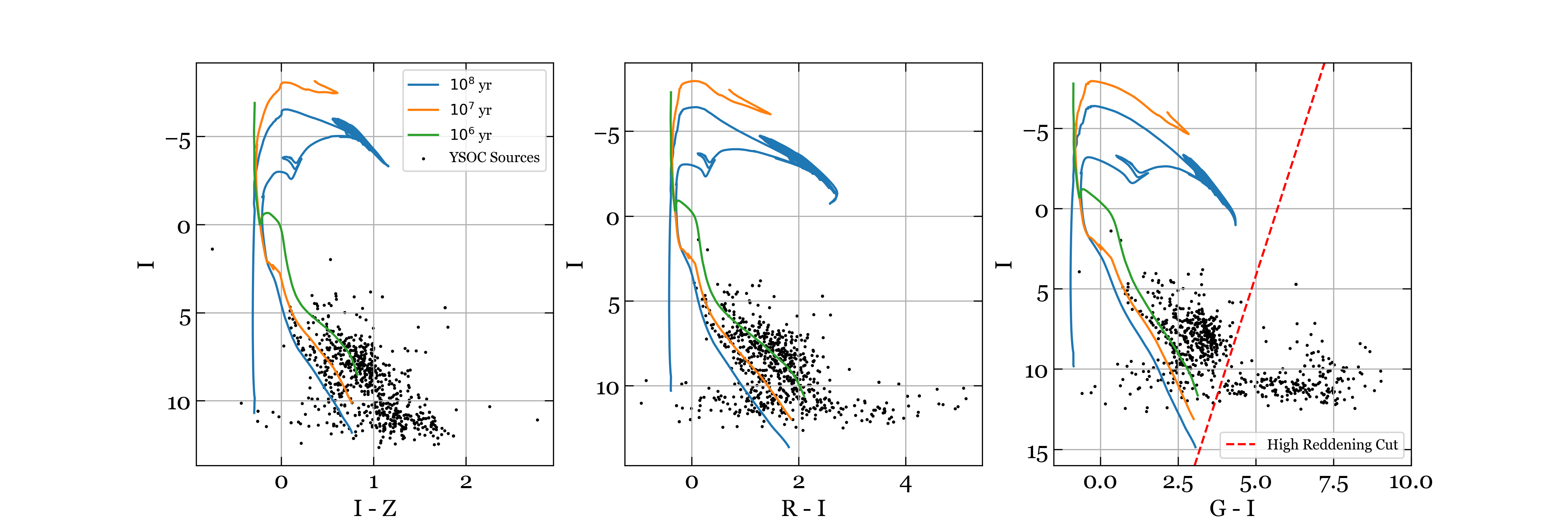}
    \caption{Sloan filter $i$ vs $i-z$, $i$ vs $r-i$, and $i$ vs $g-i$ absolute magnitude-color diagrams for YSOC stars. 
    Isochrones are from MIST \citep{Choi2016, Dotter2016, Paxton2011} with parameters  [Fe/H] = 0, v/v$_{critical}$ = 0.4, and Av = 0. 
    Red dashed line on $g-i$ diagram shows cutoff for stars we deem to be likely background due to high reddening.
    }
    \label{fig:ysocisochrone}    
\end{figure*}

 We also compare photometry for the literature sample with theoretical isochrones.  The location of the bulk of the observed distribution relative to computed pre-main sequence isochrones allows us to determine a maximum age for likely members of the cluster. Stars born from the same molecular cloud are expected to lie along sequences in color-magnitude diagrams, though especially for young embedded systems like Mon R2, reddening plays a significant role in the colors and magnitudes of our stars. 

In standard color-magnitude diagrams, pre-main sequence stars should lie above and to the right of main sequence stars at the same distance. We use photometry from PanSTARRS \citep{2020ApJS..251....7F} and Gaia \citep{2016A&A...595A...1G} to assess this for our Mon R2 sample. Typical of star-forming regions, Mon R2 has an estimated age of around 10$^6$ to 10$^7$ years \citep{herbstracine1976, carpenter2008}. In Figure \ref{fig:ysocisochrone} we confirm this age range is consistent with our sample by comparing the photometry in different color-magnitude planes to 1, 10, and 100 Myr isochrones from \cite{Dotter2016, Choi2016, Paxton2011}. The majority of our candidate member stars are in the expected location. However, some are not, and in the next sub-section we use this information to help assess membership likelihood for individual stars. 

\subsection{Establishing Membership Probability} 
\begin{figure*}[h]
    \centering
    \includegraphics[width=\linewidth]{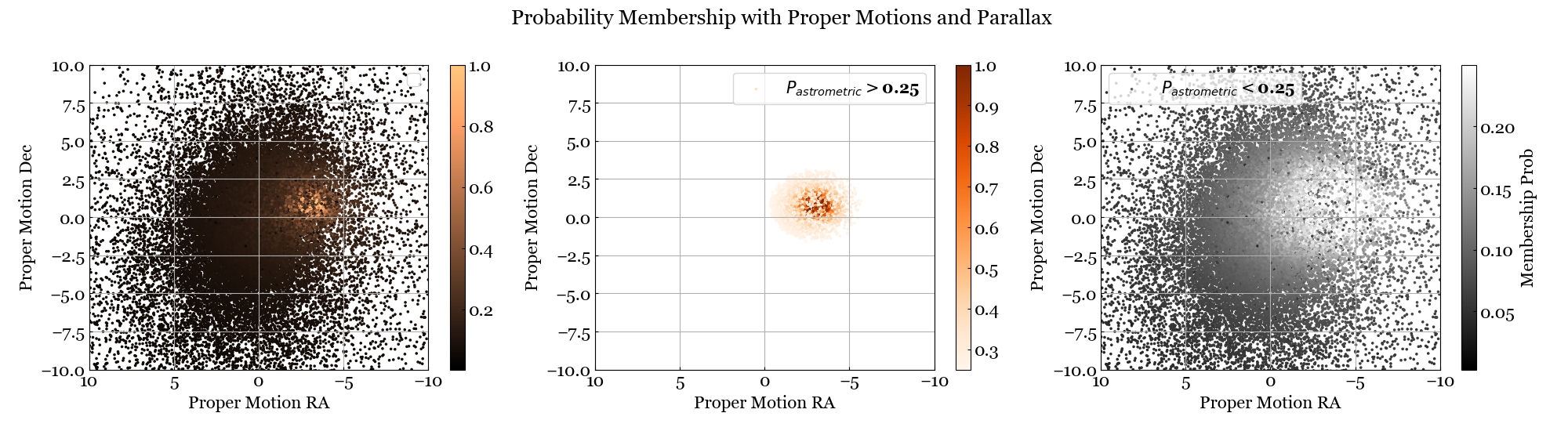}
    \caption{Left: Vector point diagram of proper motion distributions among all 
    Gaia DR3 star over the $2.5\times2.5$ deg$^2$ query area centered on Mon R2. 
    Colorbar indcates astrometric membership probability ($P_{astrom}$) as described in the text.
            Middle: Stars with astrometric membership probability $P_{astrom} > 0.25$.
            Right: Field stars excluded from the candidate member sample 
            as a result of low astrometric membership probability.}
    \label{fig:probpm}    
\end{figure*}

With the properties of probable cluster members established above, we now create a set of criteria to estimate membership probabilities that will be used to select likely cluster stars from among all stars projected on the sky in the vicinity of Mon R2. For each star, we evaluate three membership probability metrics, and calculate a total membership probability using weighted averaging. In addition to the astrometric and color-magnitude information discussed above, we also include photometric variability in the ZTF data as an additional factor in the membership probability weighting.  We label our membership metrics as probabilities, though they are not strictly such in a rigorous statistical sense, but instead a simple way of numerically ranking stars as more likely to less likely cluster members.

For our astrometric metric,  we use proper motion in RA, proper motion in DEC, and parallax, all relative to the mean cluster values quoted above. We evaluate the standard score (also called Z-score) for each, defined as $Z=\frac{X-\mu}{\sigma}$, where X is the variable of interest, $\mu$ is the variable mean, and $\sigma$ is the variable standard deviation. 
We then assign a probability of cluster membership, adopting unity for values within 1 standard deviation and 
1/Z for values more than 1 standard deviation away from the mean.
Finally, we calculate the root-sum-square of the three individual proper motions and parallax Z-scores ($Z_{pmra}, Z_{pmdec}, Z_{plx}$) and produce an astrometric probability ($P_{astrom}$) by taking the inverse of this as

\begin{equation} 
\centering
P_{astrom} = \frac{1}{\sqrt{(Z_{pmra})^2+(Z_{pmdec})^2+(Z_{plx})^2}}.
\end{equation}

Figure \ref{fig:probpm} shows the results of this criterion for the Gaia candidates. We isolate stars with $P_{astrom}>0.25$, which indicates that they should be close to the cluster mean value in at least one of the astrometric parameters. 

\begin{figure}[h]
    \centering
    \includegraphics[width=\linewidth]{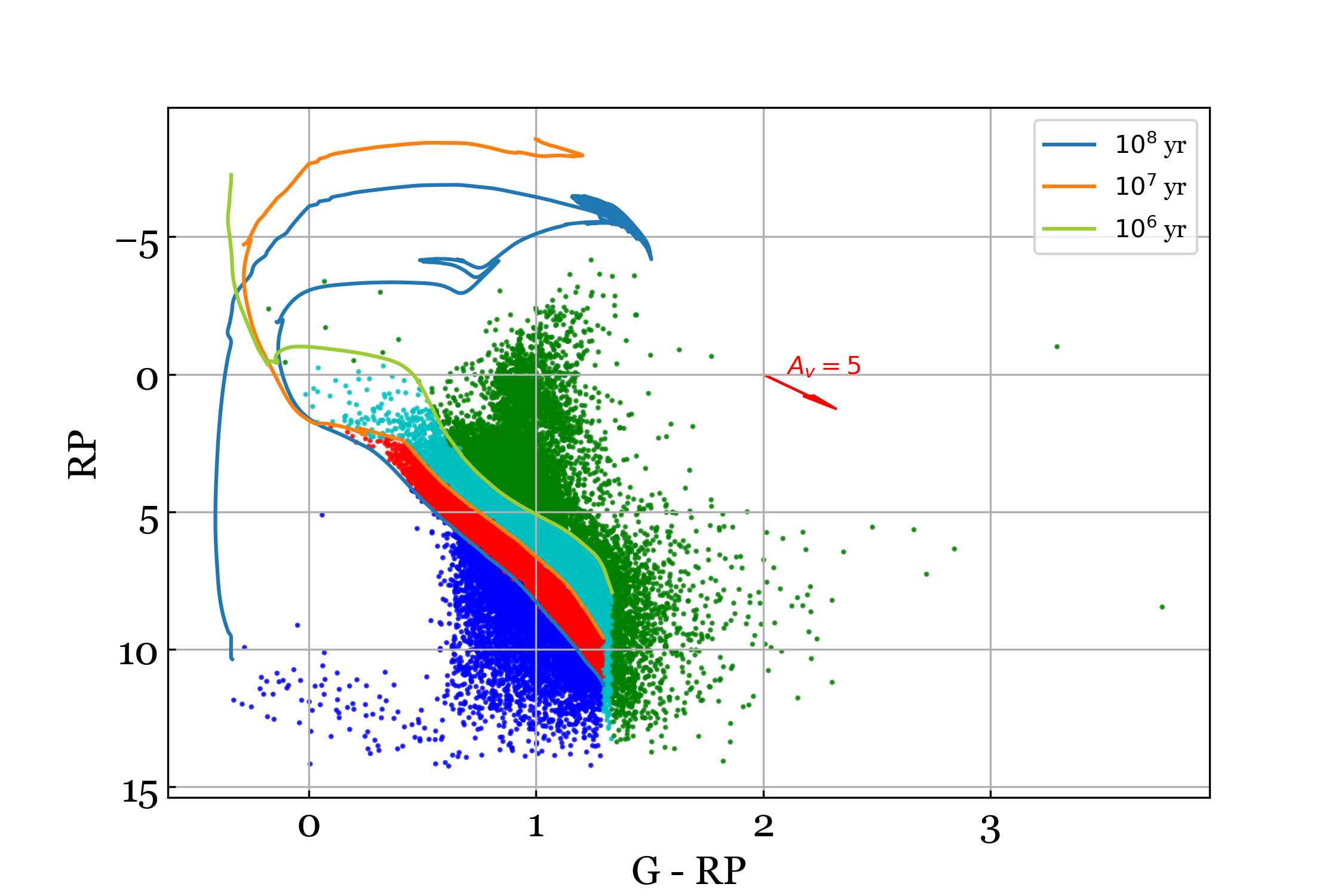}
    \caption{Isochrones compared to RP absolute magnitude and G-RP color for all stars in the initial Gaia query. Stars are colored by their location relative to the isochrones, with green points representing stars above the $10^6$ year isochrone, cyan between $10^6$ and $10^7$ year, red between $10^7$ and $10^8$ year, and blue points stars below the $10^8$ year isochrone. }
    \label{fig:initisochrone}    
\end{figure}

We also use color-magnitude diagrams based on Gaia photometry to check whether our stellar candidates are consistent with the age of the cluster and pre-main sequence stars. A G-RP vs RP color-magnitude diagram (Figure \ref{fig:initisochrone}) is used to assign a membership probability $P_{iso}$ based on the location relative to the isochrones discussed in Section \ref{subsec:establishmember}. 
In making this assessment, we consider that many of the stars are young stellar objects surrounded by gas and dust, which will affect their colors on an individual basis, and we allow for reddening up to about $A_V=5$ since the main cluster is still embedded.  We assume all stars below the main sequence at the distance of Mon R2 are field stars, in practice allowing the $10^8$ yr isochrone as the oldest potential cluster member.  Equation \ref{eq:isoprob} shows the assigned membership probabilities. 
\begin{center}
\begin{equation} \label{eq:isoprob}
P_{iso} = \begin{cases}
		0, & \text{below $10^8$ yr isochrone} \\
            0.25, & \text{between $10^7$ yr and $10^8$ yr isochrone}\\
            1, & \text{between $10^6$ yr and $10^7$ yr isochrone}\\
            0.75, & \text{above $10^6$ yr isochrone}
		 \end{cases}
   \end{equation}
\end{center}
Finally, in addition to astrometry and photometry, we assign a photometric variability $P_{var}$ metric for cluster membership. Young pre-main sequence stars are ubiquitously variable \citep{fischer2023}, and non-variable stars at the photometric precision of ZTF are highly likely to be older than 10-100 Myr, which would mean they are less likely to be members of the young Mon R2 cluster.  We assign variability probability membership values based on the measured scatter in the optical light curve ($\sigma_{mag}$)  relative to the median error (err$_{mag}$) in the data, according to Equation \ref{eq:varprob}.
\begin{center}
\begin{equation} \label{eq:varprob}
P_{var} = \begin{cases}
		0, & \text{$\sigma_{mag}$ 	$<$ err$_{mag}$} \\
            0.25, & \text{$\sigma_{mag}$ $\ge 1\times$ err$_{mag}$}\\
            0.75, & \text{$\sigma_{mag}$ $\ge 2\times$ err$_{mag}$}\\
            1, & \text{$\sigma_{mag}$ $\ge 3\times$ err$_{mag}$}
		 \end{cases}
   \end{equation}
\end{center}

We thus have astrometric, isochronal, and variability-based membership probabilities for individual stars. Some objects have all three available, while others have only two or one of these values established.

\subsection{Identifying New Candidates Members from Gaia DR3 }

\begin{figure*}[h]
    \centering
    \includegraphics[width=\linewidth]{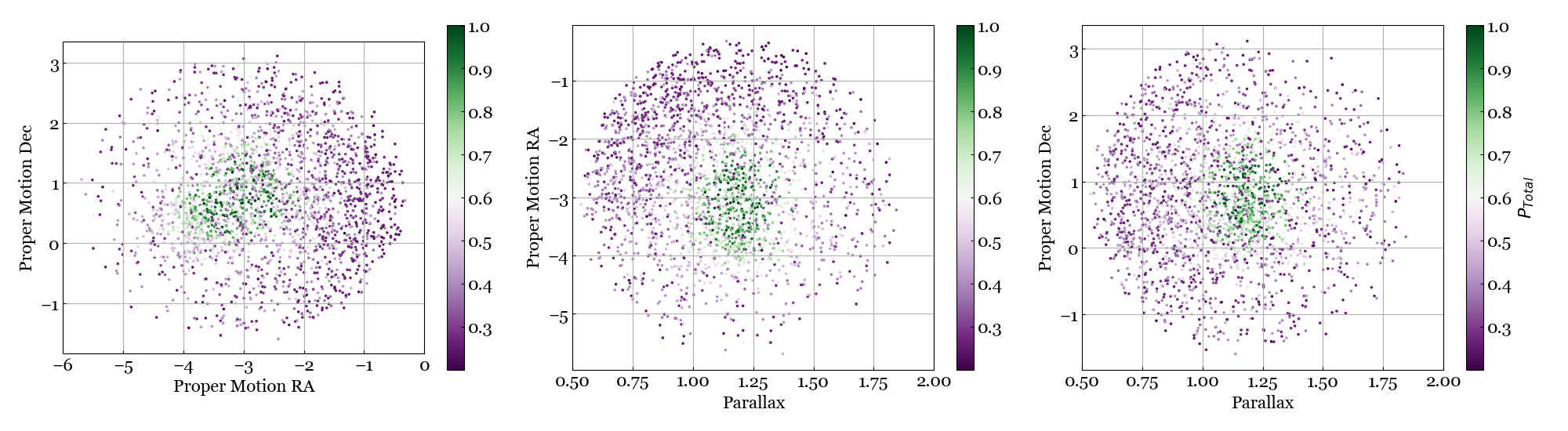}
    \caption{The distribution of total membership probability $P_{total}$, as given by the colorbar, across proper motion and parallax space for the 2551 Gaia stars with $P_{astro} >$ 0.25.  Given the weighting of the membership selection criteria, most of the highly probable members lie close to the estimated median kinematic parameters for the cluster. However, the additional criteria based on photometry relative to isochrones, and photometric variability, mean that some stars identified as candidate cluster members are further away from the kinematic centers.
    }
    \label{fig:totalprobpmplx}    
\end{figure*}

In this section we apply the membership probabilities defined and assigned above to the initial candidate list from our broad 2.5$\times$2.5 deg Gaia query of 41,493 stars around Mon R2. In order to only focus on the most likely candidates, we first require our sources to have astrometric probability $P_{astrom} > 0.25$. This cut alone leaves us with 2551 stars, for which we then calculate a weighted average total membership probability based on all three criteria (Equation \ref{eq:weitotalprob}) to find the most highly probable members. We apply weights of $w_1 = 3, w_2 = 1, w_3 = 1$, to the astrometric, isochronal, and variability probabilities in deriving the final membership probability. These weights are used, since the parameter of astrometry is composed of three separate parameters, in contrast to the remaining probabilities which are composed of one parameter each.

\begin{center}
\begin{equation} \label{eq:weitotalprob}
P_{total} = \frac{w_1*P_{astrom} + w_2*P_{iso}+w_3*P_{var}}{\sum_{i=1}^{3} w_i}
   \end{equation}
\end{center}

As noted above, not all stars have all measurements. For the sample sourced from Gaia, while all stars have astrometric values, 42 of these do not have G or RP magnitudes, and 97 do not have ZTF lightcurves. As such, for this $<$5\% of the sample, we set the weights to 0 when information is missing so as to not bias the final $P_{total}$.
Figure \ref{fig:totalprobpmplx} shows our sample of 2551 Gaia stars with $P_{astrom} > 0.25$, in proper motion and parallax space, with their final probability membership value, $P_{total}$ indicated.

\begin{figure}[h]
    \centering
    \includegraphics[width=0.45\linewidth]{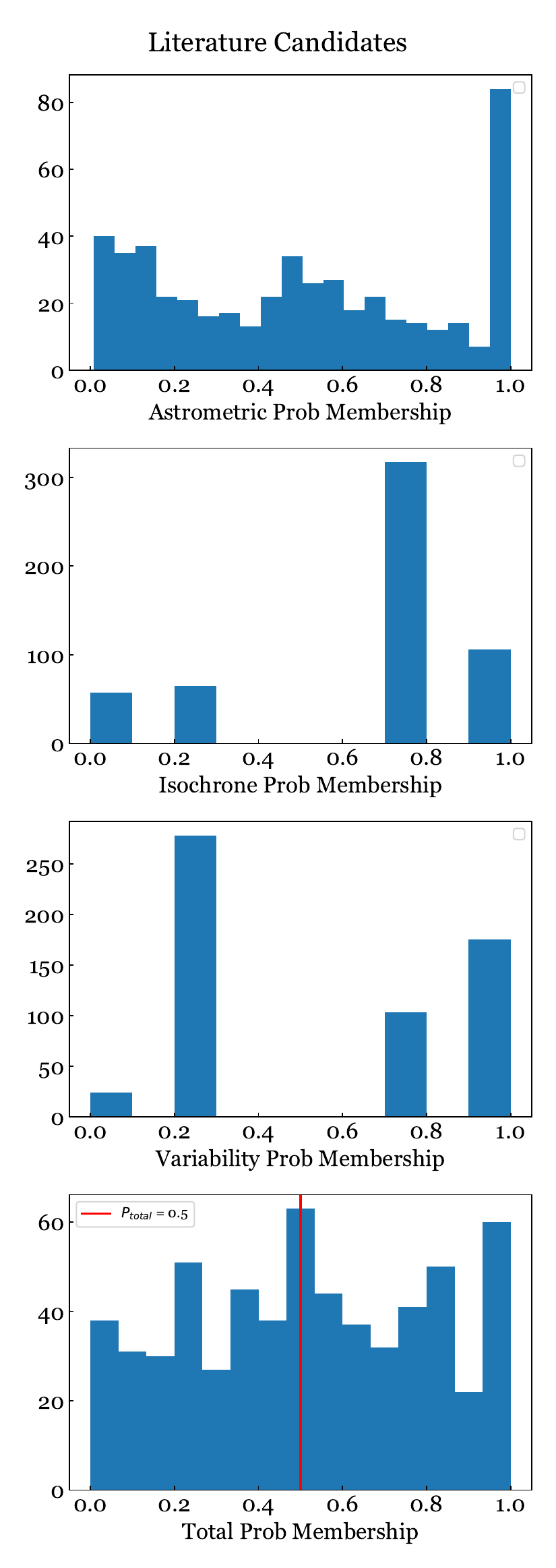}
    \includegraphics[width=0.45\linewidth]{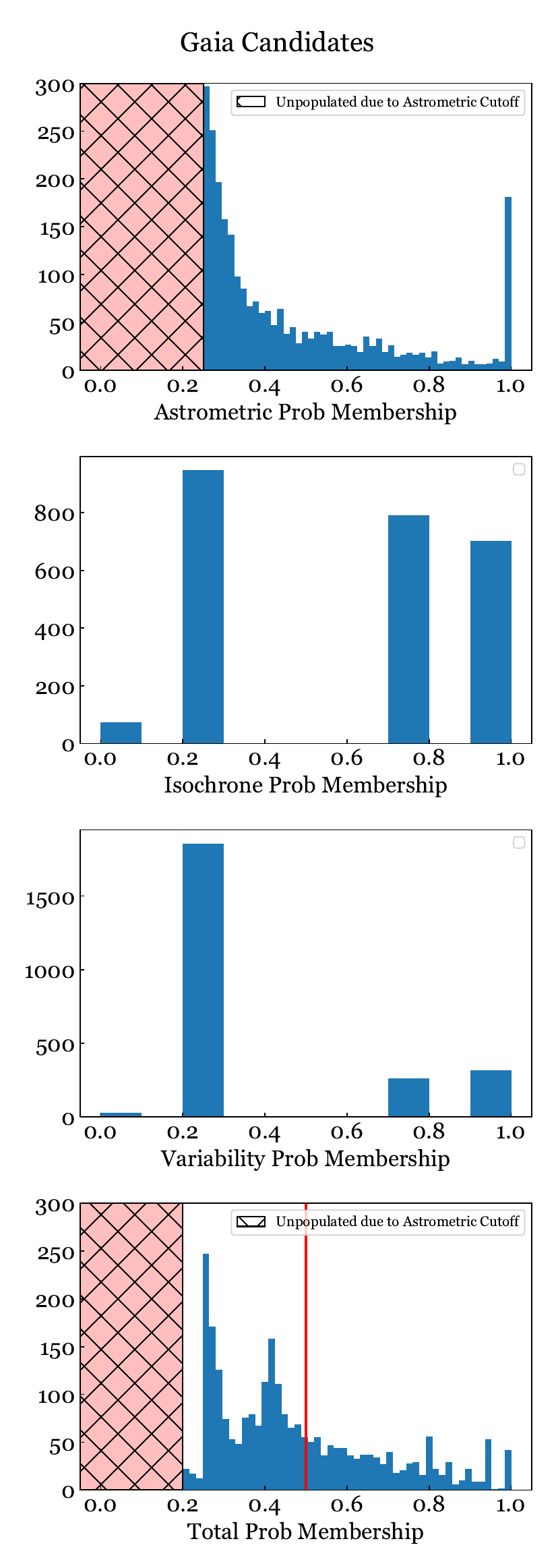}
    \caption{Membership probability distribution for each membership assessment technique shown for stars from the YSOC catalog (left) and the Gaia query (right).  The top three panels show the contributions from the astrometric, photometric isochronal, and lightcurve variability assessments, with the combined final result in the bottom panels.  We consider all stars with $P > 50\%$ in the combined histogram (red vertical line) to be likely members of Mon R2. 
    }
    \label{fig:probhist}    
\end{figure}

We select those sources with total membership probabilities $P_{total}> 0.5$ as likely members of Mon R2. Figure \ref{fig:probhist} shows the distribution of probabilities for each one of our assessed properties (astrometric, isochronal, and variability), and the combined probability used to separate members from non-members. 
The cutoff value of $0.5$ is not well-justified, especially given the fairly flat distribution in $P$. However, experimentation with different cutoff choices shows that $0.5$ represents a balance between being too 
conservative (e.g. eliminating many objects that are centrally located in clusters and have obvious YSO-like variability) and too liberal (e.g. including stars that are widely distributed across the region and not preferentially adding to the clusters).

\begin{figure*}[h]
    \centering
    \includegraphics[width=1.1\linewidth]{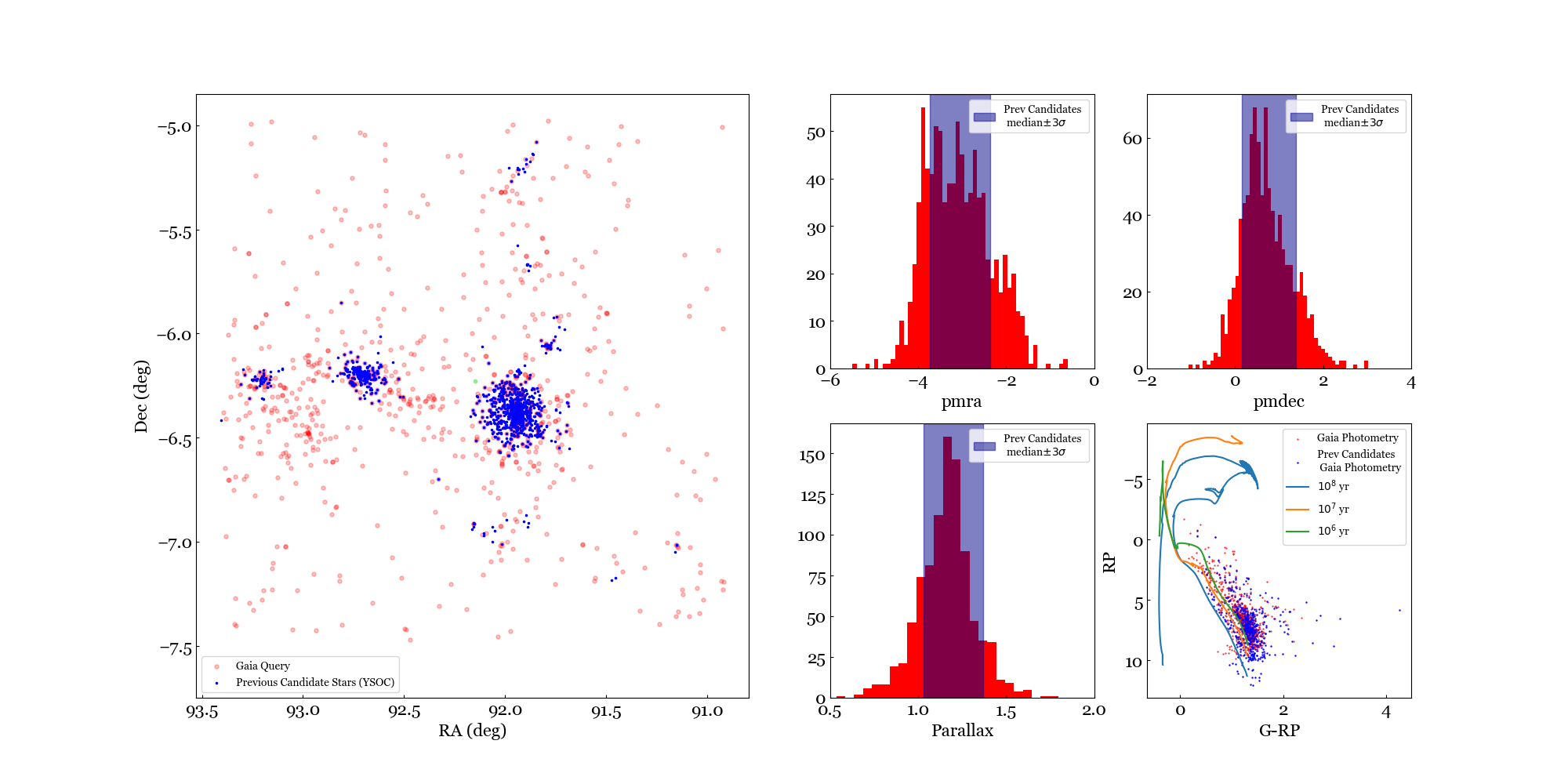}
    \caption{Same as Figure \ref{fig:initialmem}, but now showing only the 921 Gaia-selected sources that have membership probability $P_{total} > 0.5$.  Blue points are members already suspected as such from the literature, while red points are members of Mon R2 identified from Gaia. 
    }
    \label{fig:finalmem}    
\end{figure*}

The process results in 921 stars that are highly probable members. We illustrate these 921 sources in Figure \ref{fig:finalmem} relative to those previously known from the candidate member list sourced from YSOC. In \S3.5 we will assess how many of the 921 are newly identified vs. previously suspected members.


\begin{figure*}[h]
    \centering
    \includegraphics[width=\linewidth]{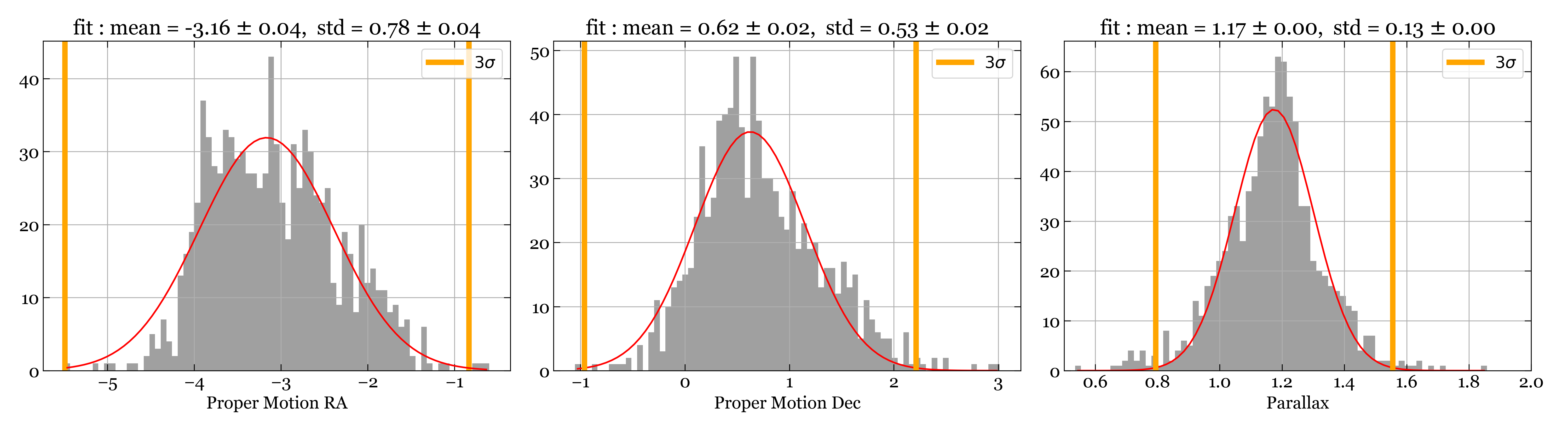}
    \caption{Gaussian fits (red lines) and 3$\sigma$ limits (orange lines) for proper motion in RA (Left), proper motion in Dec (Middle), and parallax (Right) for the final catalog of highly probable literature plus newly identified member stars. 
    }
    \label{fig:finalmem_pmplx}    
\end{figure*}

\subsection{Applying Membership Probability to YSOC Catalog Stars}

In addition to the new members we found using our Gaia query of the larger field, we can also investigate the membership probability of the stars in the YSOC catalog to see if there are any outlier candidates whose membership should be re-examined. However, many of the stars in the YSOC catalog do not have astrometric values, due to their optical faintness, but can still be likely members of the cluster. As such, we apply our membership criteria to the 496 stars with good parallax measurements and the 994 stars with no astrometry measurements. Similarly applying our process above for the stars sourced from Gaia, 
to the literature sample sourced from YSOC, we find 308 stars that are highly probable members with $P_{total}>0.5$. Figure \ref{fig:probhist} also shows the distribution of probabilities for the YSOC stars for each criterion and for the total membership probability, which can be compared to the Gaia-selected sample.
\subsection{Comparing the New Gaia Members to Previously Known Mon R2 Members }
In order to assess which of the 921 Gaia-selected Mon R2 members found in \S3.3 are newly determined as such, versus already suspected Mon R2 members from the previous literature, we cross-matched to the list of 308 highly probability members from YSOC that we reconfirmed in \S3.4. Stars with a Gaia versus YSOC coordinate separation of $<0.25\arcsec$ are taken as confirmed matches, and a small number (7) of matched sources with separations $0.25-1\arcsec$ were individually examined and are also determined to be matches. All objects in the Gaia-selected members list more than $1 \arcsec$ from any YSOC source were taken as new members.
We thus find that 270 of the 921 Gaia-selected members were previously known literature members. We have therefore newly identified 651 sources from Gaia that match the kinematics of the Mon R2 region. 

\subsection{Conclusions on the Kinematics of Mon R2}

Thus far, we have used a catalog of previously associated stars from the YSOC database in order to estimate the parameters of the Mon R2 cluster and establish a set of astrometric, isochronal, and variability membership criteria. We then used these criteria both to reconfirm the highly probable members in our YSOC list, while also identifying likely nonmembers, and to search for new highly probable members in the field discovered using the Gaia DR3 catalog. Lastly, we crossmatched our new Gaia stars with the YSOC stars to eliminate redundant sources. In the end, we are left with 959 highly probable optical cluster members, 308 from YSOC and 651 from Gaia. Table \ref{table:membershipcuts} shows the selection criteria along with the number of stars that remained at each step, leading to our final list.  Table \ref{table:members} shows our final members list, including the membership probabilities for all applied criteria. 

In this sub-section, we use our final list of highly probable members to re-estimate the kinematic parameters of Mon R2. By repeating our methods of \S3.1, where we fit one-dimensional Gaussians to the proper motions and the parallax (Figure \ref{fig:finalmem_pmplx}), we find the following means and standard deviations:

\begin{itemize}
\centering
    \item{Proper Motion RA: $-3.16 \pm 0.78$ mas/yr}  
    \item{Proper Motion Dec: $0.62 \pm 0.53$ mas/yr}
    \item{Parallax: $1.17 \pm 0.13$ mas}
\end{itemize}
with errors of $\pm0.04$ mas/yr in RA, $\pm0.02$ mas/yr in Dec, and $<0.005$ mas in parallax.
The parallax corresponds to a distance of $854 ^{+85}_{-107}$ parsecs.

We note that our elimination of some previously claimed Mon R2 candidate members, and our addition of new, previously unassociated, stars has left the
standard deviations on the proper motions and parallax the same or even smaller
than reported above for the literature sample. 

\begin{table}[t]
\caption{Sample size as a function of membership cuts for the literature sources (YSOC) and newly identified sources (Gaia)\label{table:membershipcuts}}
\begin{tabular}{c|c|c}
\hline
\hline
Criterion & YSOC Stars & Gaia Stars  \\ 
\hline
Initial List & 1690 & 107557  \\
In Gaia DR3 & 696 & 107557  \\
Quality Parallax & 496 & 41493  \\
$P_{astro} > 0.25$ & \dots & 2551  \\
$P_{total} > 0.5$ & 308 & 921\\
New Candidates & \dots & 651 \\
\hline
\end{tabular}
\end{table}

\section{Optical Variability Methods} \label{sec:opticalvar}

In this section, we classify the optical variability of the stars we have determined are members of Mon R2, using light curves from ZTF.  We use both variability metrics and a periodicity search to select a subset of the cluster members list for further analysis. 
We then classify these stars based on the Q-M classification established by \cite{cody2014}.

We work only with 889 stars from the 959 star membership list, based on requiring the ZTF light curve to have $>$25 observations so that we can reliably calculate the variability metrics.  Furthermore, while we used variability above as one of the criteria in assessing new member candidates, with a relative weighting of only 20\%, there are some candidates with high membership probabilities that turn out to be not formally variable, and for which we do not pursue the Q-M classification. 
\subsection{Variability Metrics}\label{sec:var_metric}
\begin{figure}
    \centering
    \includegraphics[width=1.1\linewidth]{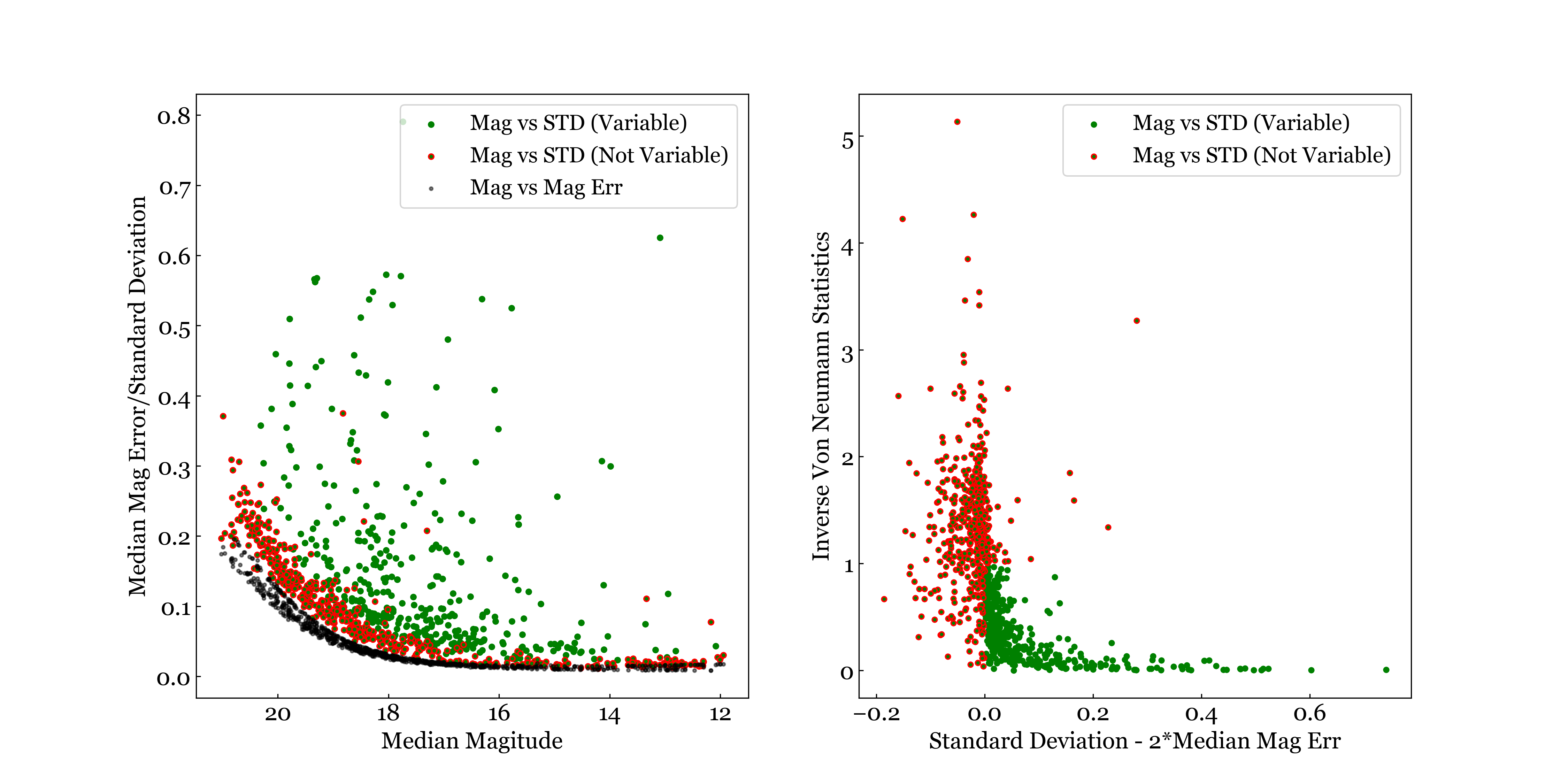}
    \caption{Left: 
    Black points indicate median magnitude, $m_{med}$ versus median error, $err_{mag}$, in the ZTF data.
    Red and green points indicate median magnitude versus lightcurve standard deviation, $\sigma$ with red outline indicating stars that
    do not meet a threshold of standard deviation, $\sigma_{mag}$, divided by median magnitude error, $2\times err_{mag}$, larger than two, and green those that do. Right: excess variability, plotted as standard deviation ($\sigma_{mag}$) minus twice the median magnitude error ($2\times err_{mag}$) versus inverse von-Neumann statistic. Variable stars are selected as those having inverse von-Neumann $<$1 and $\sigma_{mag} > 2\times err_{mag}$}
    \label{fig:varstats}    
\end{figure}

Using the ZTF r-band magnitudes and magnitude errors,
we calculated ten commonly used metrics to quantify variability behavior.  These metrics are: median ($m_{med}$), mean ($m_{avg}$), root mean square ($m_{rms}$), standard deviation ($m_\sigma$ or $\sigma_{mag}$ hereafter), kurtosis ($m_{kurt}$), skew ($m_{skew}$), median absolute deviation (MAD), inverse von neumann statistic ($\eta$), and Chi-Square ($\chi^2$). 

Additionally, the median of the magnitude error over the time series (err$_{mag}$) 
is calculated to identify baseline noise for the variability assessment. 
Our metrics are those established by the ZTF Source Classification Project \citep{2021MNRAS.505.2954C}. 

For some light curves, there are outlier points that are inconsistent with the rest of the lightcurve. In most cases, these are spurious noise that would bias the calculation of many of our metrics. To identify such contamination, we use the kurtosis metric or a measure of the “tailedness” of the data distribution, selecting all light curves with kurtosis $>4$ for examination. We check whether the high kurtosis is caused by a few outlier points, or more systematic outliers that are likely astrophysical. For example, large amplitude variables with short timescale changes would have high kurtosis values. In total, there are 25 
stars with high kurtosis due to outlier points; these sources have their variability metrics recalculated using only observations within three standard deviations of the median brightness. For the stars with high kurtosis that is likely due to astrophysical outliers, we keep all data points of the light curve. 

We classify objects as variable stars if both of the following criteria are met. (1) the standard deviation ($\sigma_{mag}$) of the light curve is larger than twice the median of the magnitude error ($err_{mag}$), and (2) the inverse von-neumann statistic is less than unity. Figure \ref{fig:varstats} shows the distribution of median magnitude vs median magnitude error (black) and standard deviation (green) for all our stars. Stars with a red outline are not considered variable as they do not fit both our criteria. 

\subsection{Period Search} \label{sec:per_metric}

\begin{figure}
    \centering
     \includegraphics[width=\linewidth]{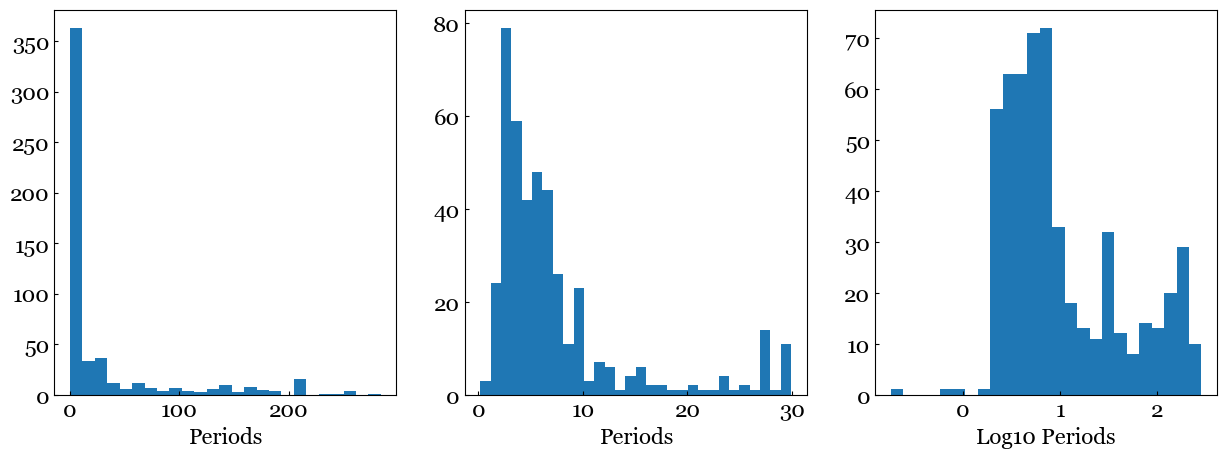}
     \caption{Distribution of most significant Lomb-Scargle peaks for all stars.  Left panel shows the full distribution; center panel a zoom-in on the lower range of values that are most likely to be true rotation periods rather than a mix of rotation and other timescales associated with young star variability; right panel shows a logarithmic format of the full period distribution.}
    \label{fig:period}    
\end{figure}

We used the \textsc{astropy LombScargle} periodogram package to find and analyze periodic signals in our ZTF light curves. We set our initial period search to find periodicity between 2 and 250 days. This range avoids the cadences from Earth’s rotation (daily) and motion around the sun (annual), but special attention does need to be paid to possible confusion of astrophysical timescales with the lunar orbit cadence ($\approx$30 days). The resulting periodograms are used to find the most likely period and its significance, as quantified by the standard false alarm probability (FAP). 

Of the 889 stars for which we performed the period search, we found 466
significant (FAP $<0.01$) periods. For these stars, we visually examined all observed and phased light curves, and determined whether the periods were justified, or if an expanded period search below 2 days or above 250 days would be informative. For stars where none of the plausible periods seemed appropriate, but the star met the variability criteria set in \S4.1, we manually label the source as Aperiodic and assign it the timescale having the highest power. 

We ended up eliminating 59 sources with approximately monthly or 250+ day timescales that seem unreliable.  We also readjusted the periods of approximately 32 stars, some to ranges $<2$ or $>250$ days when such periods provided better phased light curves than those from our initial search.  Figure \ref{fig:period} shows the distribution of significant/believable periods. The majority of stars in our sample have a period $<10$ days, suggestive of typical stellar rotation timescales for a young population. All sources with significant periods are considered for our variable star sample, even if the object does not meet both variability criteria articulated in subsection \ref{sec:var_metric}. 
Ultimately, we retain those added based on their significant periodicity only if they also have a low value of the quasi-periodicity metric $Q$, as described below.

\subsection{Q and M Metrics and Variability Classification}\label{subsec:qmclassify}

\begin{figure*}[h]
    \centering
    \includegraphics[width=\linewidth]{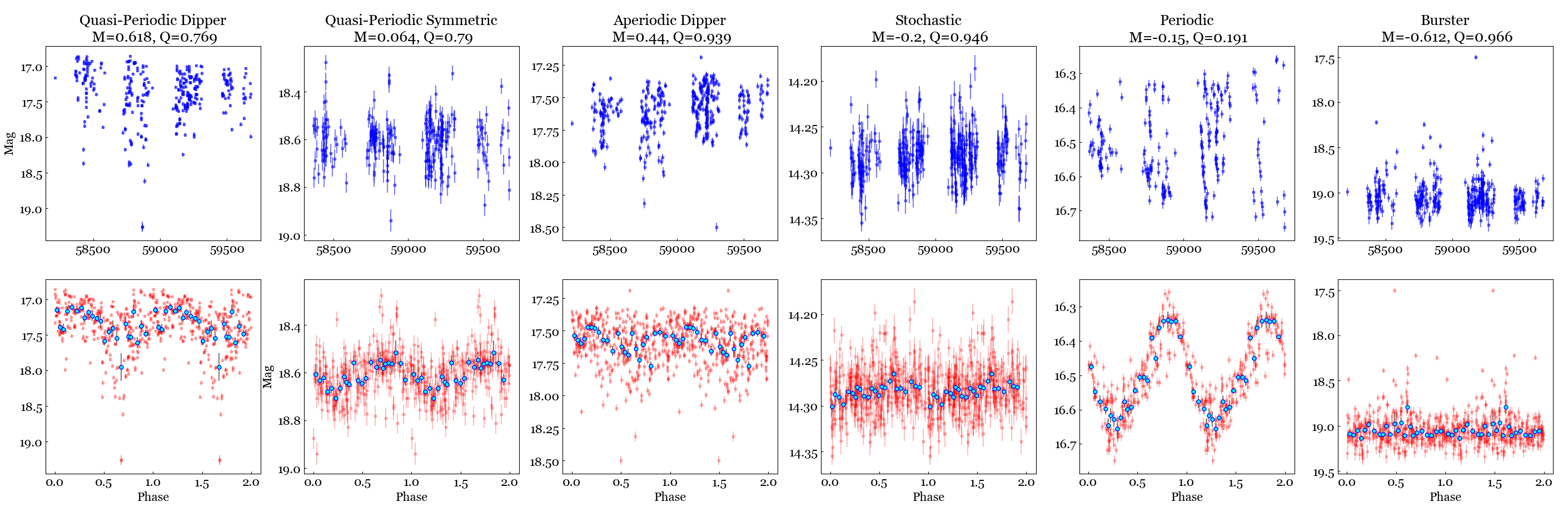}
    \caption{Example sources with different $Q-M$ classification, as labeled. 
    Top: Observed light curves for the entire four seasons of ZTF data; error bars are shown. 
    Bottom: Phased light curves using the highest peak in a Lomb-Scargle periodogram (red points) with binned median magnitude (blue points); error bars are shown. Q and M values are indicated.}
    \label{fig:qmclass_ex}    
\end{figure*}

For Q-M classification, we require the star to either be a statistically significant photometric variable, defined above based on the lightcurve having a standard deviation more than twice the median error, and an inverse von Neumann statistic $\nu <$ 1,
or the lightcurve having a significant period, with false alarm probability of less than 1\%.

If a star has a period that is deemed not significant, and it does not meet the variable star criteria, it is removed from our list for Q-M classification. In total, among 889 
stars, we end up with 470 stars that are variable, significantly periodic, or both. 

For each of these stars, we used the Q and M metrics first codified in \cite{cody2014}, and also applied in \cite{cody2018} and \cite{cody2022} to young star lightcurves that were obtained from spacecraft (CoRoT and K2, respectively).
Application of this same technique to ZTF data by \cite{hillenbrand2022} was taken as proof of concept for the present study.  $Q-M$ classification of young star light curves results in a parameter space of periodicity to aperiodicity ($Q$), and fading to bursting lightcurves ($M$). 

The Flux Asymmetry, or M metric, measures whether a light curve has brightened or dimmed over time, and is defined as:
\begin{equation}\label{eq:M_metric}
    M = \frac{\langle m_{10\%}\rangle - m_{med}}{\sigma_{mag}},
\end{equation}
where 
$\langle m_{10\%}\rangle$ is the is the mean magnitude of the top 90\% and bottom 10\% percentile
$m_{med}$  is the median magnitude, and
$\sigma_{mag}$ is the standard deviation of the magnitude. 

The Quasi-Periodicity, or Q metric, is a measure of whether a light curve contains purely periodic signals, quasi-periodic signals, or aperiodic signals. To measure $Q$, we apply the formula: 
\begin{equation}\label{eq:Q_metric}
    Q =\frac{\sigma^2_{resid} - \sigma^2_{photo}}{\sigma^2_{mag} - \sigma^2_{photo}},
\end{equation}
by first determining the dominant timescale identified in the Lomb-Scargle period finding process described above, and subtracting this purported periodic signal from the observed lightcurve
to produce the $\sigma_{resid}$ term.  The $\sigma_{phot}$ term is the mean photometric error in the lightcurve and $\sigma_{mag}$ is the standard deviation of the magnitude measurements. 

The initial run produced a small number of stars with Q values outside the range of 0 to 1. Upon visual inspection, it was determined that these sources had either a small number of measurements, or did not meet the formal variability criteria and had been included based only on their low-FAP periods. As these periods were close to the lunar cadence,  all such sources were eliminated from the $Q-M$ analysis. 

Stars with M values $>$ 0.25, have predominantly dimming light curves and are considered dippers, while stars with M values $<$ -0.25 are predominantly brightening and considered bursters, while M values between -0.25 to 0.25 have relatively symmetric lightcurves and are considered symmetric. Stars with low $Q$ values have a more strictly periodic signal, with most of the lightcurve subtracting out well from the smoothed periodic signal while a star with a $Q$ value of 1 would be very stochastic or aperiodic, having a large signal even after subtraction of the periodic signal. We considered stars with $Q<$ 0.45 as periodic, $0.45 < Q < 0.82$ as quasi-periodic, and $Q > 0.82$ as aperiodic.   These boundaries were validated by visual inspection. We note that very few significant periods (FAP $< 1\%$) are found at $Q > 0.48$,
and stars with such high $Q$ should be taken to have ``timescales" rather than true periods.

With $M$ and $Q$ calculated for each star, we assigned the $Q-M$ categories as in \cite{cody2014}: Periodic (P), Quasi-Periodic Symmetric (QPS), Quasi-Periodic Dipper (QPD), Aperiodic Dipper (APD), Stochastic (S) -- which corresponds to an aperiodic-symmetric signal -- and Burster (B), a class that includes both aperiodic and quasi-periodic bursters.  
We then manually inspected all of the phased lightcurves to validate the dominant timescale reported from the Lomb-Scargle analysis.
If we determined in the manual inspection that the chosen period was inaccurate, we looked for other (usually shorter) plausible periodogram peaks that phased well, and reassigned the $Q$ value, and possibly the classification between aperiodic, quasi-periodic, and periodic.  If none were found, we removed the $Q$ value calculated for the star and assigned it a classification in one of the aperiodic categories.  

\begin{figure*}[h]
    \centering
    \includegraphics[width=\linewidth]{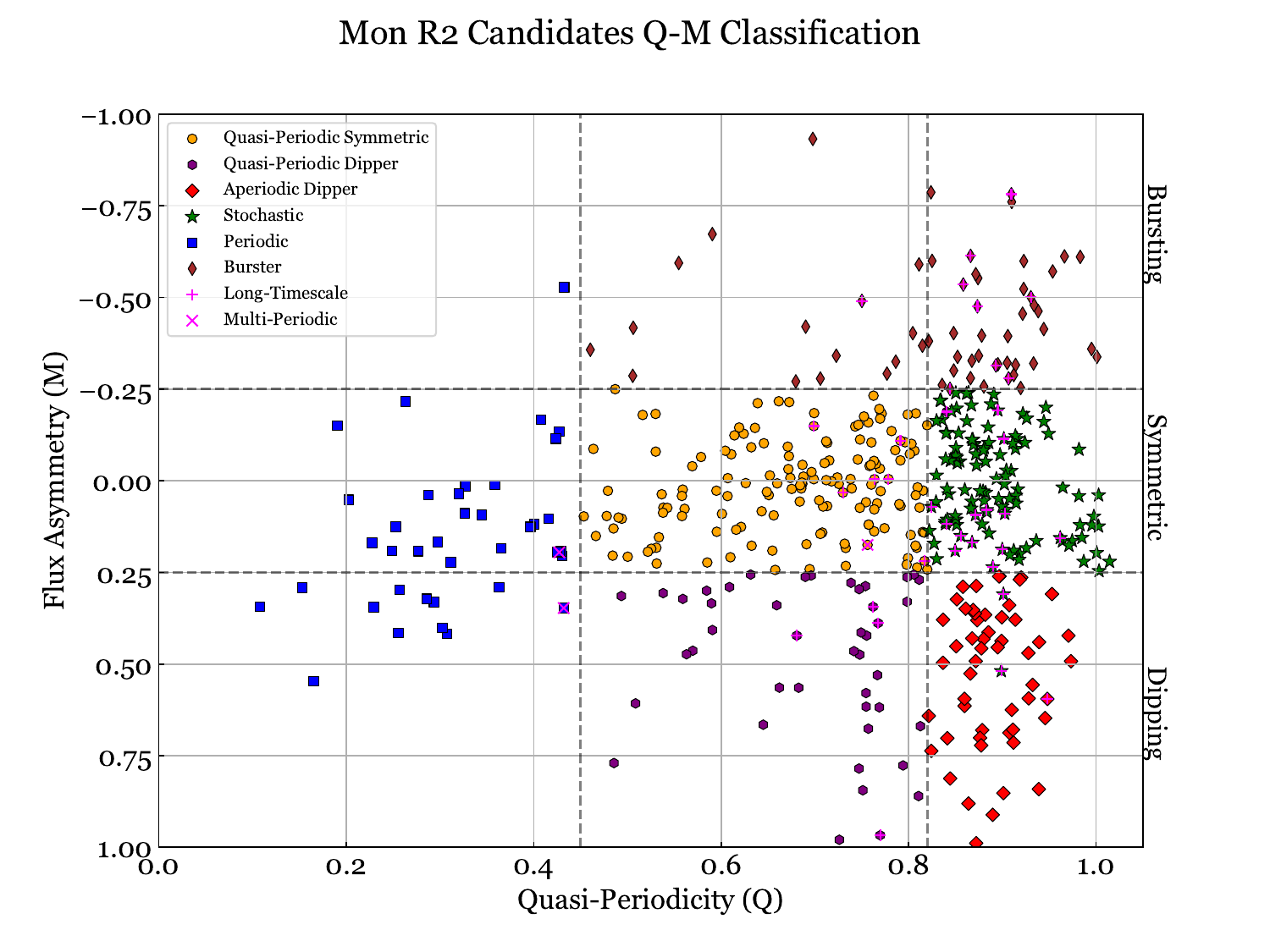}
    \caption{Flux Asymmetry ($M$) versus Quasi-Periodicity ($Q$) plotted for our sample of 470 
    variable member sources. Primary variability classes are shown by the marker indicated in the legend. Sources with a secondary classification (multi-periodic or long timescale) are marked in addition to their primary class. Dotted lines show the limits in $Q$ and $M$ for the differing classes, with $Q=0.45$ and $Q=0.82$ dividing the periodic from quasi-periodic and quasi-periodic from stochastic sources, and $M=-0.25$ and $M=+0.25$ dividing the burster from symmetric and symmetric from dipper sources.}
    \label{fig:qmclassify}    
\end{figure*}

Figure \ref{fig:qmclass_ex} shows an example star for each of the classification groups, with both the entire light curve and the phase-folded version illustrated. 
The distribution of our final $Q-M$ classifications for Mon R2 is shown in Figure \ref{fig:qmclassify} 
and the complete list of sources and their $Q-M$ classes is given in Table \ref{table:vari_class}.
Table \ref{table:qmclass_summary} provides a summary of the different variable star classes.

\begin{table*}[t]
\centering
\caption{Summary of QM classification for Variable and Periodic Stars in Mon R2 \label{table:qmclass_summary}}
\begin{tabular}{c|c|c}
\hline
\hline
Class & All Stars & Long Timescale \\
\hline
Aperiodic Dipper (APD) & 55 & 2 \\
Burster (B) & 59 & 9  \\
Periodic (P) & 39 & 0  \\
Quasi-Periodic Dipper (QPD) & 46 & 5 \\
Quasi-Periodic Symmetric (QPS) & 153 & 7 \\
Stochastic (S) & 118 & 14 \\
\hline
TOTAL & 470 
& 37\\
\hline
\end{tabular}
\end{table*}

\subsection{Stars with Additional Timescales}

\begin{figure}[h]
    \centering
    \includegraphics[width=\linewidth]{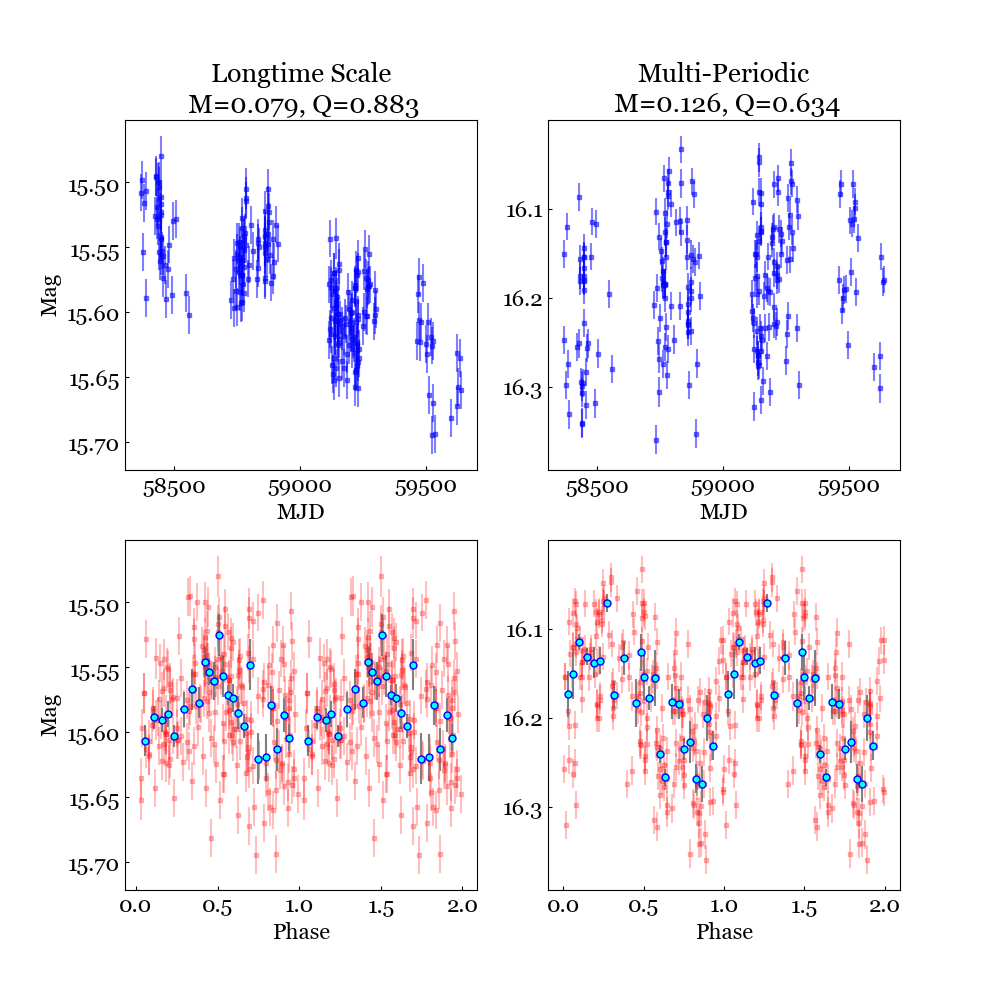}
    \caption{Same as Figure \ref{fig:qmclass_ex}, but example sources with secondary classification of long timescale (L) or multi-periodic (MP).}
    \label{fig:qmsec_ex}    
\end{figure}

In our periodograms, the most significant (tallest) peak is used to calculate the $Q$ metric.  However, some lightcurves show additional timescales that we describe in this section.
We flagged stars that we considered to have long-timescale or truly multi-periodic behavior, and added a secondary class to the primary $Q-M$ classification indicating multi-periodic (MP) or long timescale (L) behavior. 

We identify potential multi-periodic signals by flagging 
stars with more than one peak above the 1\% significance power threshold. To validate such candidates, we extract all significant periods from the periodogram and manually inspect each phase-folded light curve. If there are multiple periods that phase well and are not aliases of one another, we give the light curve a secondary classification of Multi-Periodic (MP). In total, we have only two stars with the secondary classification of MP. 

To identify long timescale stars, we searched for light curves that have systematic changes in brightness over a significant portion or even the entirety of the four-year lightcurve time span. The stars with a secondary classification of long timescale (L) are all named as such through manual inspection. 

Figure \ref{fig:qmsec_ex} shows one example of each type of secondary classification.

\subsection{Correlations with the Q-M Classes}
In this section, we consider how well the $Q-M$ variability classes align with standard statistical metrics of variability.  
We also investigate the variability behavior of the population with respect to physical location in the Mon R2 region. 

For each source, we first simplified its variability classification to just the $M$ metric designation as a burster, symmetric, or dipper.
Then, we examine the variability metric distributions across the simplified classes. 

We find that dippers have on average a lower inverse von-Neumann statistic than symmetrics and bursters, making them more readily identified as variables. This is further evident from examining the variability level as measured by standard deviation over median magnitude error, where we find the most variable stars are also most likely to be dippers, while symmetrics make up the majority of the least variable stars, likely due to their relatively low amplitudes. 
The burster sources tend to have dimmer light curves, on average, than the symmetric and dipper sources. 
It may be that these stars are picked out as variables based on the burst behavior, whereas comparably faint stars with symmetric or dipping variability would be harder to identify as variables based on small-amplitude changes relative to their median brightness especially given that the noise levels are increasingly higher for fainter stars.

We also note that there are no dippers or bursters with $Q<0.4$; in other words, all of the truly periodic sources at low $Q$ have relatively symmetric lightcurves.
This is consistent with other studies, and indicative that the mechanisms causing the aperiodicity in the lightcurves also
cause the skew, measured here as elevated values of the $|M|$ metric.  In these cases, the presumed culprit is circumstellar material.

Another trend in the variable star population is that dipper sources appear to be concentrated toward the central cluster of Mon R2,
and toward the clusters found to the east and to the north. 
Under the assumption that dipper behavior is likely associated with circumstellar material,
which is known to be correlated with stellar youth, 
the clustering of the dipper population supports another common hypothesis that clustering is related to stellar youth.

\subsection{Large Amplitude Variables}

While examining our dataset, we identified several tens of stars with light curves exhibiting substantial changes in magnitude. Some of these stars are well known and have been studied in detail based on their previously identified erratic behavior. A few such examples include: V899 Mon \citep{Ninan2015, Park2021}, V1818 Ori \citep{Chiang2015}, and [CMD97]-1031 \citep{Jiang2022}. We labeled these and other sources as Large-Amplitude Variables (LAVs) by manually inspecting all lightcurves 
exhibiting peak-to-peak magnitude changes of at least one magnitude, and 
having standard deviation greater than three times the median magnitude error. 

In total, we identify 52 stars that are LAVs. Their full light curves are illustrated in Appendix \ref{appen:LAV_lc},
segregated by the simplified burster, symmetric, and dipper classes according to their $M$ values. There are 8 bursters, 12 symmetrics, and 32 dippers. 
\section{Discussion}\label{sec:discussion}

In this paper, we have presented new information on the optically visible stellar population of Mon R2.
First, we have used Gaia DR3 to assign membership probabilities to stars known in the literature
as potential members.  Second, we have used ZTF to study the variability properties of these members. Our discussion centers around the topics of clustering in the Mon R2 region, and comparison to other studies of the Mon R2 population, including both kinematics and variability.

\subsection{Clustering in the Mon R2 Region}
\begin{figure*}[h]
    \centering
    \includegraphics[width=\linewidth]{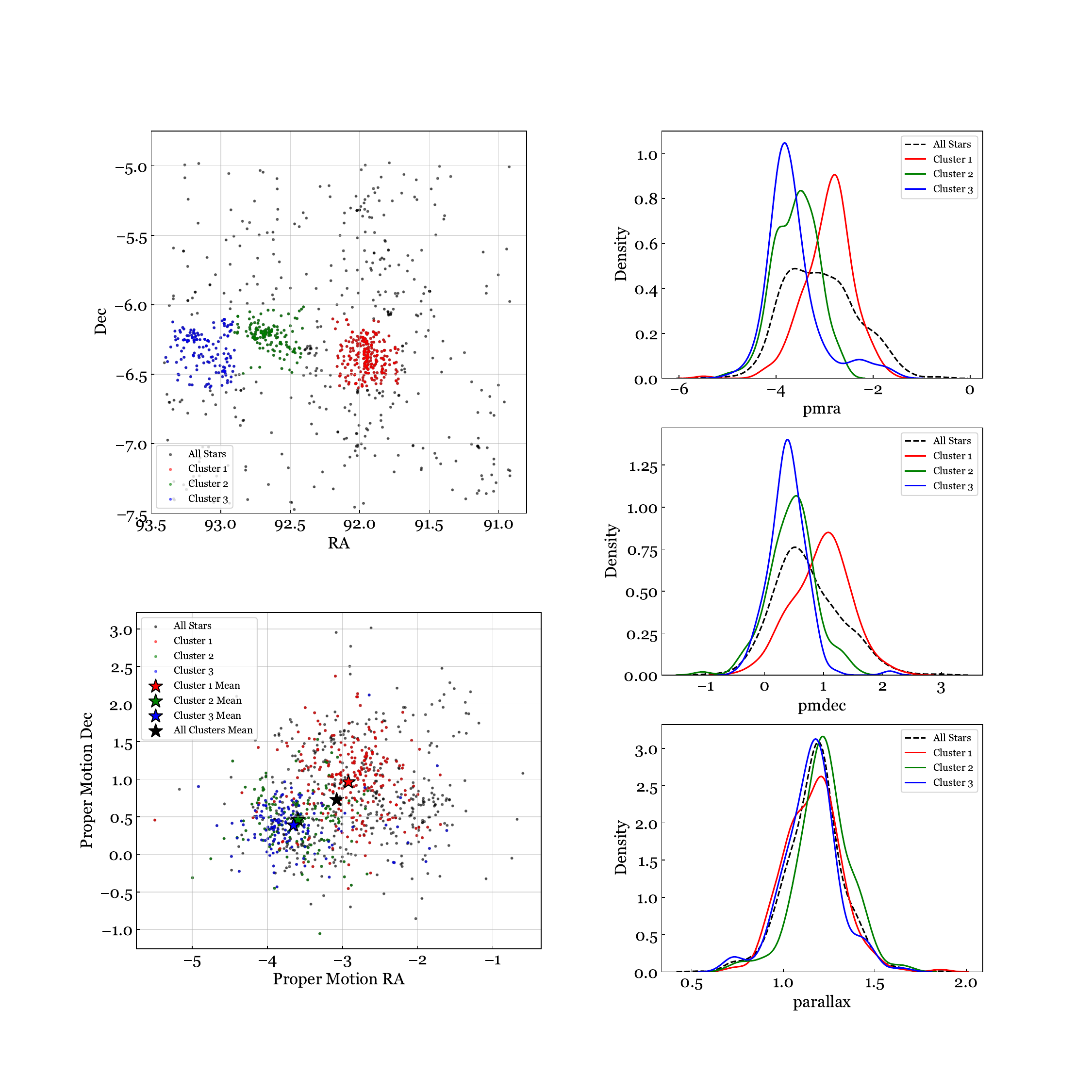}
    \caption{Left:  RA-Dec and PMRA-PMDEC spaces for Mon R2 member stars that are not clustered (black) and associated with the three sub-cluster regions (red, green, blue).  Sub-clusters are chosen by grouping within equal area 0.5 x 0.5 deg regions, centered at coordinates provided in Table \ref{table:subclusters}
    Right: Density distribution for proper motions and parallax of the three clusters (red, green, blue curves) compared with those for all member stars (black dotted curves).
    }
    \label{fig:clusters}    
\end{figure*}

Using our final member list, we examined the kinematic properties of three previously known sub-clusters in Mon R2. 
Cluster 1 is Mon R2 main, Cluster 2 comprises GGD 12-15, and Cluster 3 is the GGD 16-17 region. 
We defined three adjacent 0.5 x 0.5 deg rectangular regions 
and examined the proper motions and parallax distributions of the stars within each group. 
Figure \ref{fig:clusters} and Table \ref{table:subclusters} shows locations of the sub-clusters and the kinematic results.

The parallax distribution between the three groups is the same. However, the proper motions of stars in Cluster 1 are more positive in each direction than the motions measured for stars in Cluster 2 and Cluster 3. These values overlap within one standard deviation, which is expected given that we selected our members to have similar astrometric values. However, we do see that the majority of stars in subcluster 1 are moving on average in a different direction from stars in subclusters 2 and 3. This indicates both shared and distinct sub-motions of our final member stars, whose origins and futures could inform star formation histories and evolution in this region. 

We also looked at Gaia photometry for the subcluster stars, examining their relative distributions in the RP vs G-RP magnitude-color diagram. All three sub-clusters overlap significantly, meaning that they have similar ages.  The more distributed population, that is, stars that are not part of any of the three sub-clusters, have somewhat lower luminosity. According to the isochrones, these non-clustered stars in the region are an older population than the more clustered stars. 
We explored the variability class population in each subcluster and found slightly more dippers proportionally in the central cluster (cluster 1, with 27.5\%) than the other cluster regions (cluster 2 with 17.3\%, and cluster 3 with 20\%). This could be due to the more embedded nature of the central cluster region, as explained before, which causes more dust and material to block the light from the young star.

\begin{table*}[t]
\centering
\caption{{Table of Positions and Astrometric values of stars in each subcluster and for all stars in our sample}}\label{table:subclusters}
\begin{tabular}{c|c|c|c|c}
\hline
\hline
Name & RA, Dec & PMRA & PMDec & Parallax \\
 & deg, deg & mas/yr & mas/yr & mas \\
\hline
Cluster 1 & 92.65, -6.35 & -2.91 $\pm$ 0.55 & 0.96 $\pm$0 .31 & 1.16 $\pm$ 0.15 \\
Cluster 2 & 92.65, -6.25  & -3.59 $\pm$ 0.43 & 0.45 $\pm$ 0.39 & 1.21 $\pm$ 0.14 \\
Cluster 3 & 93.15, -6.35 & -3.65 $\pm$ 0.55 & 0.38 $\pm$ 0.31 & 1.14 $\pm$ 0.14 \\
All Regions & - &  -3.16 $\pm$ 0.78 & 0.62 $\pm$ 0.53 & 1.17 $\pm$ 0.13 \\
\hline
\end{tabular}
\end{table*}

\subsection{Comparison of Results with Previous Works}
\subsubsection{Comparison of Kinematic Results}

In this section, we compare our kinematic results to two other works that have conducted kinematic studies on populations in Mon R2. 

\cite{Kuhn2019} used a sample of 97 candidate member stars identified in previous surveys and located towards the core of Mon R2 (the center of our Cluster 1). These authors calculated a weighted median value for proper motion RA = $-2.91\pm0.11$ mas/yr, proper motion Dec. = $1.05\pm 0.18$ mas,/yr, and parallax = $1.06\pm 0.04$ mas, corresponding to distance = 948$^{+42}_{-38}$ pc. 

More recently, \cite{Orcajo2023} examined stars over the central 26\arcmin $\times$ 26\arcmin region in the Mon R2 region using the Las Cumbres Observatory Global Telescope Network (LCOGT). They matched stars they found in LCOGT to Gaia EDR3 stars, and selected a large rectangular region in PMRA-PMDec space defined by 
(-2.75, 1.15) mas/yr $\pm$ 1.25 mas/yr to study photometric variability. For parallax, they performed a Gaussian fit to the distribution of stars within their proper motion box, and found values within 0.888 to 1.42 mas as likely members, corresponding to distance = $825 \pm 51$ pc for the cluster. 

The region examined in our work is much larger (2.5 deg $\times$ 2.5 deg) than the works mentioned above, both of which encompassed smaller regions centered on our Cluster 1. Although our membership methods differ substantially, and our parent sample size and is much larger, we find that our measurements of proper motion and parallax are consistent and contained within one standard deviation of the values found in these previous works. Comparing only our Cluster 1 values, we find even better agreement between our kinematic values and those found previously.

\subsubsection{Comparison of Variability Results}
Regarding variability properties of stars in Mon R2, we can compare our results to two other studies that have also used ZTF data to study the variability of stars in young star-forming regions using the \cite{cody2014} $Q-M$ statistics.  Caveats to any such comparison include the different membership selection criteria used among the studies, as well as the different distances and extinctions of the clusters,  which could mean that somewhat different populations are being probed with ZTF photometry. 

We find that our population of stars in Mon R2 has similar proportional distribution among the $Q-M$ classes 
as found in the ZTF study by \cite{hillenbrand2022} of the North America and Pelican (NAP) Nebulae region.
However, there are some differences in results between our study and that of the Perseus (Per) Molecular Cloud by \cite{wang2023}, also using optical photometry from ZTF. 

Specifically, our population of bursters (12.5\%) and aperiodic dippers (11.7\%) is comparable to those found in the NAP region (B: 13.9\% / APD: 9.6\%). For Quasi-Periodic Dippers, Mon R2  (9.7\%) has a more similar population to Per (7.5\%), than the larger population of this variable type found in the NAP region (19.2 \%).  We find that there is a much lower population in Mon R2 of purely periodic variability, and a much greater population of stochastic variability compared to NAP and Per. 
Among all three regions and studies, Quasi-Periodic Symmetric dominates the variability classification. This is likely due to the optical photometry picking up days- to weeks-long variability that is modulated on the stellar rotation timescale, whether originating from the stellar surface itself, or in circumstellar regions near the co-rotation radius in the disk. 

Next, we can compare our variability results to a recent search for variable stars also conducted in Mon R2 by \cite{Orcajo2023}. 
These authors
present data spanning 23 days over which 1500 photometric observations were aquired. 
Our ZTF data, by contrast, covers a much longer baseline (4 years) but with fewer samples (33-570 photometric observations per star).  
We crossmatched our 470 member variables to the 136 sources \cite{Orcajo2023} designated as having ``rotation-modulated variability", and compare the classifications and periods. We find some apparent differences in astrometry, necessitating use of a 5 arcsec radius to match sources. We found 59 stars in common between the two studies, all of which were visually examined to ensure the same astrophysical source was being compared.  Among the 59 matches, only 39 of the reported periods are within 1 day of each other.  A handful of stars in the ``rotation-modulated variability" list have reported periods of one day, which we had disregarded among our sample in order to avoid the diurnal signal.  Further investigation showed that all of the sources with mismatched periods have very high Q values, making them aperiodic sources with ``timescales", rather than truly rotationally-modulated behavior which is expected to be sinusoidal.  For the period-mismatched stars, we find that using the period reported by \cite{Orcajo2023} results in a more aperiodic (higher $Q$) signal from our ZTF data, and as such we believe that our periods (or timescales) are more correct for our dataset. Out of the 59 matched stars, only 6 are classified by us as purely periodic, or the equivalent of a rotation-modulated variable, while
another 29 are defined in our analysis as "Quasi-Periodic Symmetric", which is related to stellar rotation but likely measuring phenomena in the co-rotating circumstellar environment. 
\section{Conclusion and Summary}
To summarize, we gathered 1690 candidate Mon R2 members that had been identified in previous literature and examined the proper motion and parallax properties of the young stellar population. We derived their mean kinematic values. We then created a weighted membership scheme for testing new candidate members that included astrometry, location in color-magnitude diagrams, and optical variability. 

From 107557 field stars queried in Gaia DR3 over a 2.5 deg $\times$ 2.5 deg region centered at Mon R2, we thus selected 921 highly probable member stars having the expected kinematic and photometric behavior of the young stellar population of Mon R2.  We then reassessed the membership for our 1690 YSOC literature candidates using the same techniques, and crossmatched the highly probable members with the stars sourced from Gaia.  
Including both the previously proposed members sourced from the literature, and the newly identified members from the broader field, we confirmed a total of 959 Mon R2 members in this study.
Due to the optical selection of our sources, our membership list is incomplete and does not include more embedded members, mainly those fainter than $r>21$.

Our compiled Mon R2 member sample has the following kinematic properties:
\begin{itemize}
    \item{Proper Motion RA: $-3.16 \pm 0.78$ mas/yr}  
    \item{Proper Motion Dec: $0.62 \pm 0.53$ mas/yr}
    \item{Parallax: $1.17 \pm 0.13$ mas}
    \end{itemize}
Furthermore, our study has found for the first time that the previously known sub-clusters in the region have distinct kinematic properties.
We studied the proper motions among three sub-clusters in the region, the central (hub) cluster of Mon R2, and two additional clusters located to the west of it. 
The central cluster kinematics we measure here agree well with values reported in two previous works that study this region. However, we additionally find different proper motion directions for the two western clusters (Cluster 2 and Cluster 3) than for the central cluster (Cluster 1), which impacts the kinematic values we measure for the Mon R2 region overall.

For our compiled member list for Mon R2, we then collected optical r-band lightcurves from ZTF. We identified 470 sources with sufficiently high photometric sampling in ZTF that could be classified as variable based on excess scatter ($\sigma_{mag} > 2\times err_{mag}$) or significant periodicity (Lomb-Scargle False Alarm Probability $<$ 1\%) sources. We then classified the optical variability using $Q-M$ classification which divides stars by their lightcurve repeatability and flux asymmetry. 

We find that the majority of the variability occurs at timescales of less than 10 days, and is dominated by Quasi-Periodic Symmetric (32.5\%) and Stochastic (25.1\%) variability phenomena. We also find that our population contains 11.7\% (55) Aperiodic Dippers, 12.5\% (59) Bursters (Aperiodic and Quasi-Periodic), 8.2\% (39) Periodics, and 9.7\% (46) Quasi-Periodic Dippers. In addition, we find by visual inspection that 7.8\% (37) of our stars have a longtime scale variability, while just 0.4\% (2) exhibit possible multi-periodic variability. 
We also find that dipper sources (M $>$ 0.25) are on average more variable and brighter than their burster counterparts. Finally, we identified a list of 52 large-amplitude variable stars whose fluxes go through greater than 1 magnitude of change over the span of the entire light curve. 
Through this work, we have expanded and updated the stellar census in Mon R2 through the use of kinematics from Gaia DR3 and optical photometry from Gaia and ZTF. We also examined the optical variability in Mon R2 in detail. The results of our study can be used to further understanding of the evolution of young stellar populations.

\section*{acknowledgments}
We thank the Caltech SURF/WAVE program for financial support of this work. We acknowledge previous work by Adric Reidel in compiling the Mon R2 literature and for its ingestion into YSOC, and Michael Kuhn and Luisa Rebull for general conversation regarding Mon R2. 

\vspace{5mm}
\facilities{Gaia, PO:1.2m(ZTF)} 
\appendix

\section{Lightcurves of Large Amplitude Variables}\label{appen:LAV_lc}

In this appendix, we illustrate the lightcurves of several large-amplitude variables in the Mon R2 region.  They are identified based on large values of the standard deviation, kurtosis, or $M$ metric.

\begin{figure*}[h]
    \centering
    \includegraphics[width=0.7\linewidth]{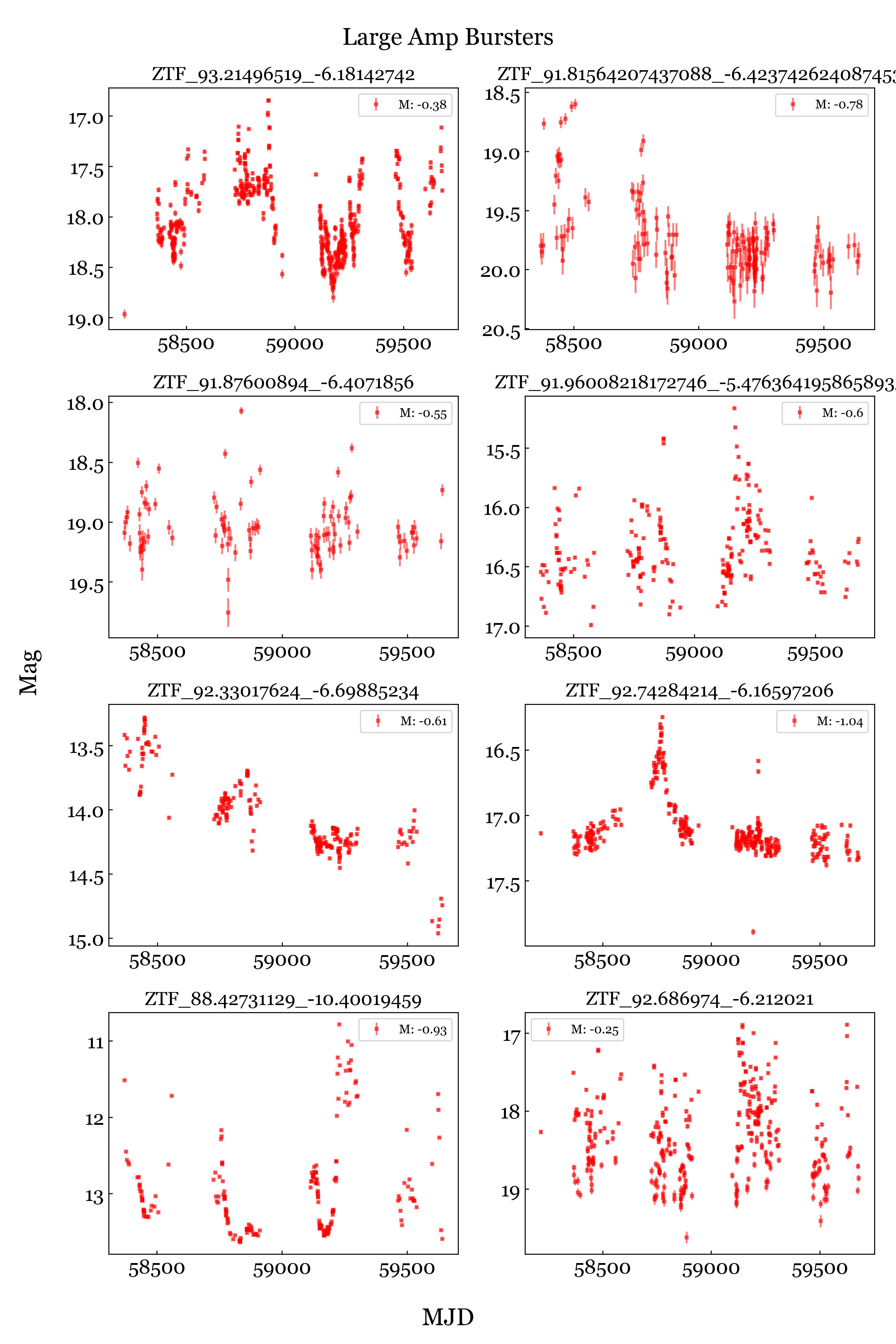}
    \caption{Light Curves with Error for all Large Amplitude Variables that have Burster Classification (M $<$ -0.25)
    ZTF\textunderscore92.33017624\textunderscore-6.69885234 or V899 Mon has been studied in more detail in \cite {Ninan2015, Park2021}
    ZTF\textunderscore88.42731129\textunderscore-10.40019459 or V1818 Ori has been studied in more detail in \cite{Chiang2015}}
    \label{fig:LAV_burster}    
\end{figure*}

\begin{figure*}[h]
    \centering
    \includegraphics[width=0.9\linewidth]{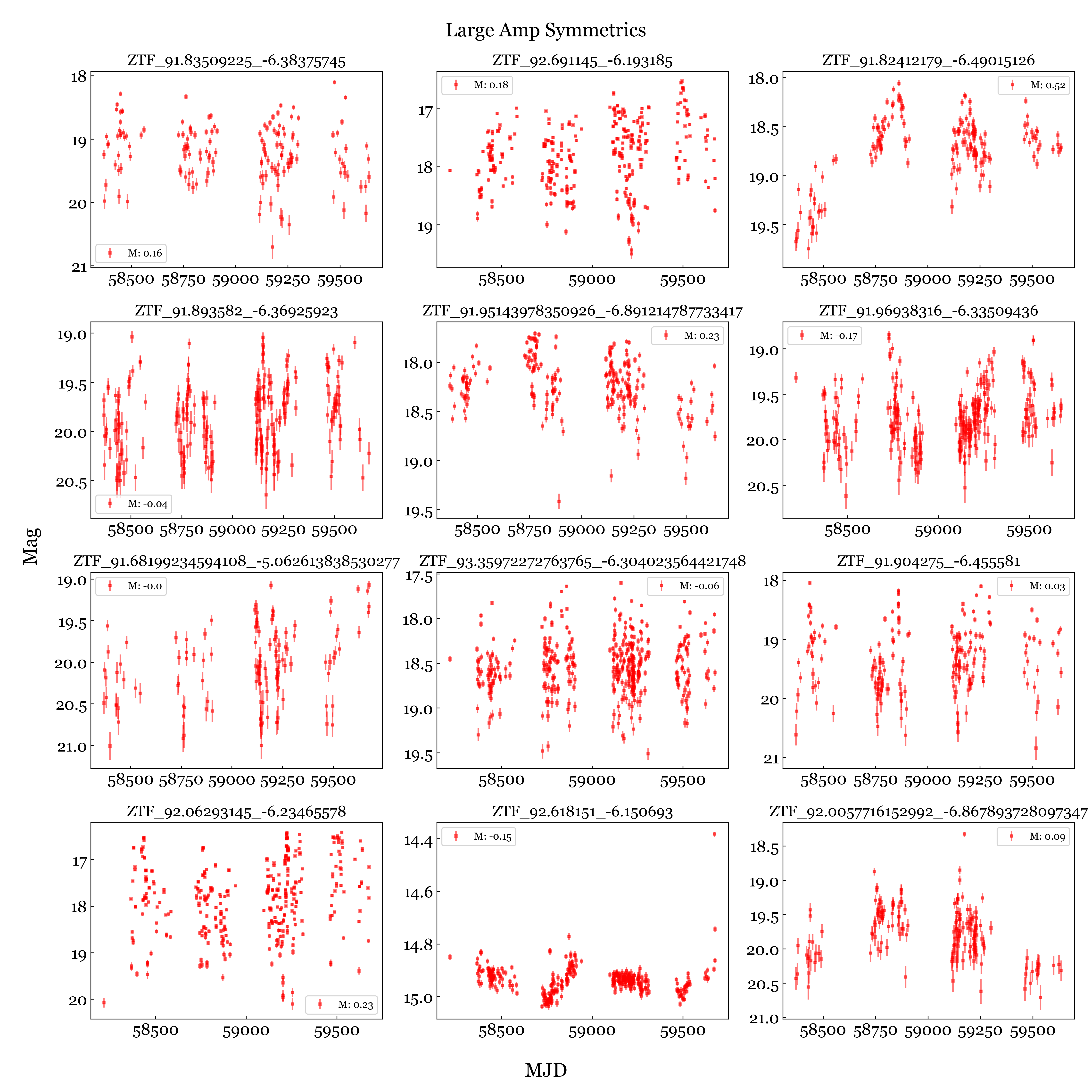}
    \caption{Light Curves with Error for all Large Amplitude Variables that have Symmetric Classification. ZTF\textunderscore92.618151\textunderscore-6.150693 appears to have brightened recently, but over many epochs has shown both rising and dimming in brightness. Recent ZTF photometry has shown the magnitude has dimmed very quickly shortly after its rise. }
    \label{fig:LAV_symm}    
\end{figure*}
\begin{figure*}[h]
    \centering
    \includegraphics[width=0.9\linewidth, height=1.25\textwidth]{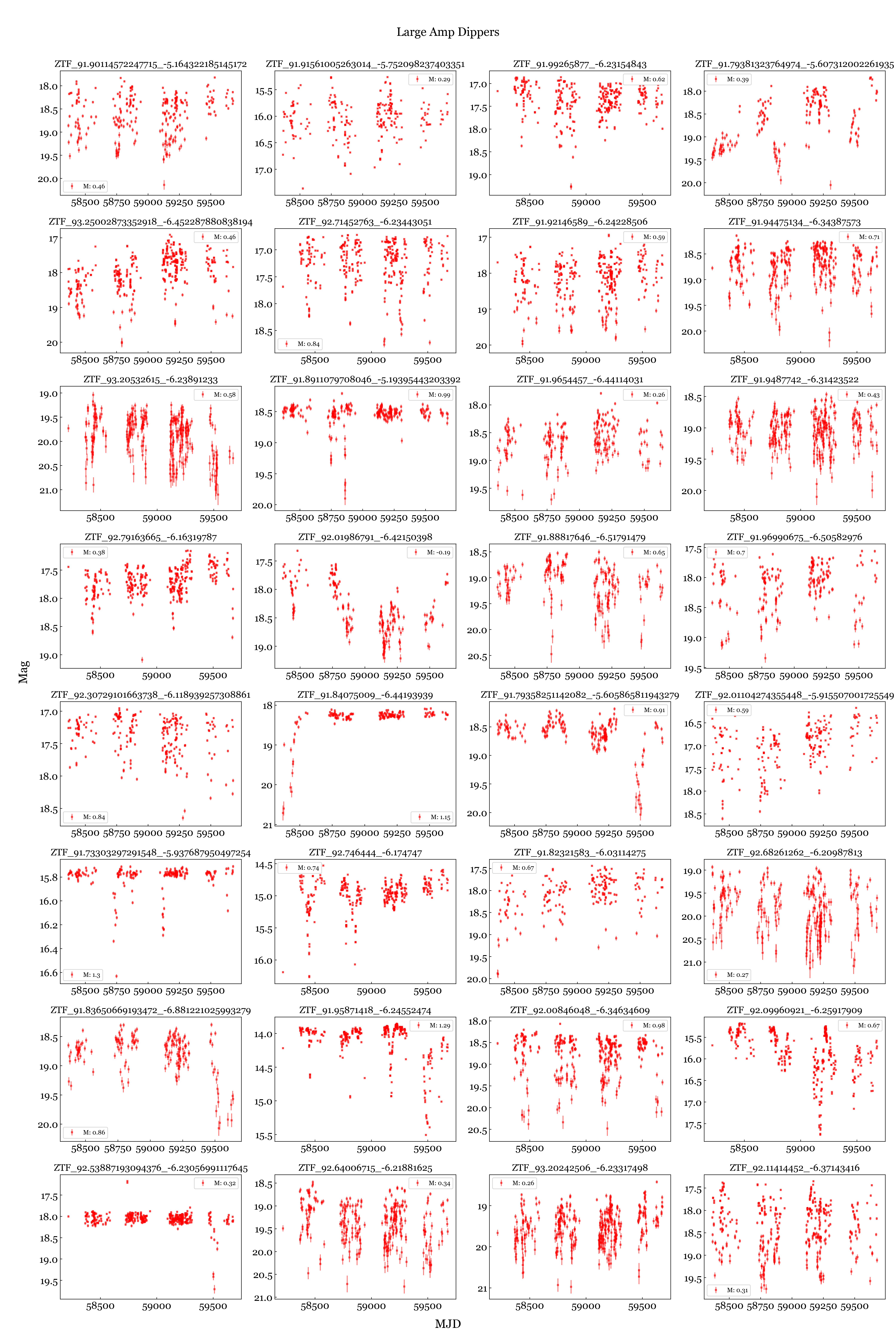}
    \caption{Light Curves with Error for all Large Amplitude Variables that have Dipper Classification (M $>$ 0.25). 
    ZTF\textunderscore91.84075009\textunderscore-6.44193939 has been studied in \cite{Jiang2022}}
    \label{fig:LAV_dipper}    
\end{figure*}
\clearpage

\begin{table}
    \caption{Table of all final members from Literature (YSOC) and new query (Gaia) \label{table:members}}
    \centering
    
    \begin{tabular}{lrrrrrrrrrrrr}
    \hline
    \hline
DESIGNATION & \multicolumn{1}{l}{RA} & \multicolumn{1}{l}{Dec} & \multicolumn{1}{l}{PMRA} & \multicolumn{1}{l}{PMDEC} & \multicolumn{1}{l}{PLX} & \multicolumn{1}{l}{$P_{ISO}$} & \multicolumn{1}{l}{$P_{ASTRO}$} & \multicolumn{1}{l}{$P_{VAR}$} & \multicolumn{1}{l}{$P_{TOTAL}$} \\ 
\\ & \multicolumn{1}{l}{degrees} & \multicolumn{1}{l}{degrees} & \multicolumn{1}{l}{mas/yr} & \multicolumn{1}{l}{mas/yr} & \multicolumn{1}{l}{mas} &  &  & &  \\ 
         \hline
GDR3 3018234532930198656 & 92.15735 & -6.91543 & -2.95 & 1.16 & 0.83 & 0.75 & 0.44 & 1.00 & 0.62 \\
GDR3 3019739936147841280 & 92.33018 & -6.69885 & -2.66 & 0.89 & 1.28 &  0.75 & 1.00 & 1.00 & 0.95 \\ GDR3 3018444294837100928 & 91.98359 & -6.58545 & -3.09 & 2.37 & 1.29 & 0.75 & 0.37 & 1.00 & 0.57 \\
2MASS J06075548-0635075 & 91.98121 & -6.58543 & -3.30 & 1.73 & 1.05 & 0.75 & 0.54 & 0.25 & 0.53 \\
GDR3 3018444741513732992 & 92.02765 & -6.56769 & -2.74 & 0.41 & 1.05 & 0.75 & 0.86 & 0.25 & 0.72
    \end{tabular}
    \tablecomments{DESIGNATION - object identifier from either previous literature obtained from YSOC or Gaia DR3 number \\ 
$P_{ISO}$ - probability membership value based on location on Gaia photometry isochrones \\ 
$P_{ASTRO}$ - probability membership value based on astrometry value in comparison to cluster's median values \\
$P_{VAR}$ - probability membership value based on variability of object \\
$P_{TOTAL}$ - total membership probability calculated with weighted average \\ 
For more detail see Section \ref{sec:membership}}
\end{table}

\begin{table}
    \fontsize{5}{7}
    \selectfont
    \caption{Table of all variable stars and their variability metrics \label{table:vari_class}}
    \centering
    \begin{tabular}{ccccccccccccccccc}
    \hline
    \hline
        DESIGNATION & RA & DEC &  $m_{med}$ &  $m_{avg}$ &  err$_{mag}$ &  $\sigma_{mag}$ &  $m_{skew}$ &  $\chi^2$ &  $\eta$ & $m_{kurt}$ &  Period &  FAP &  M &  Q  & Primary & Secondary \\
        & \multicolumn{1}{c}{degrees} & \multicolumn{1}{c}{degrees} & \multicolumn{1}{c}{} & \multicolumn{1}{c}{} & & &  & &  &  & Days &  &  &  &  & \\
        \hline
    GDR3 3019956226404877696 & 92.38052 & -6.31246 & 19.07 & 19.04 & 0.05 &  0.19 & -2.74 & 30.45 & 0.55 & Low & 4.179 & 1.67E-01 & -0.61 & 0.97 & B & \\
    GDR1 3019966190728318080 & 91.95285 & -6.28380 & 18.03 & 18.05 & 0.03 & 0.10 & 1.20 & 9.77 &  0.44 & Low & 86.778 & 2.00E-01 & 0.60 & 0.95 & APD & L \\
    GDR3 3019780652437838592 & 93.12625 & -6.30538 & 16.40 & 16.40 & 0.02 & 0.05 & 0.54 & 10.48 & 0.19 & Low & 8.282 & 1.98E-21 & 0.35 & 0.43 & P & MP \\
    GDR3 3019733854474171776 & 93.15389 & -6.30515 & 15.45 & 15.45 & 0.01 & 0.03 & -0.04 & 4.89 & 0.93 & Low & 2.894 & 3.28E-05 & -0.06 & 0.79 & QPS & \\
    GDR3 3019955715303860864 & 92.30989 & -6.31210 & 18.59 & 18.58 & 0.04 & 0.26 & -0.09 & 56.78 & 0.09 & Low & 2.857 & 2.65E-03 & -0.09 & 0.98 & S & 
   \end{tabular}
\tablecomments{$m_{med}$ - median magnitude of ZTF light curve\\
$m_{avg}$ - average magnitude of ZTF light curve\\ 
err$_{mag}$ - median of the magnitude error \\ 
$\sigma_{mag}$ - standard deviation of the magnitude\\
$m_{skew}$ - skew or measure of symmetry of magnitude\\ 
$\chi^2$ - Chi-Square of the magnitude \\ 
$\eta$ - inverse von-neumann statistic \\
$m_{kurt}$ - kurtosis or "tailedness" of light curve, high kurtosis sources have light curve data points cut to only within 3 standard deviations of median (eliminate errors)\\
Period - highest peak on Lomb-Scargle Periodogram, verified by visual inspection \\
FAP - False Alarm Probability calculated by Lomb-Scargle Peridogram \\
M - M-metric, or measure of Flux Asymmetry, calculated using variability metrics, Equation \ref{eq:M_metric}\\ 
Q - Q-metric, or measure of Quad-Periodicity, calculated using variability and Period metrics, Equation \ref{eq:Q_metric}\\ 
Primary - primary QM classification based on Q and M value, See Table \ref{table:qmclass_summary} for all names\\ 
Secondary - secondary classification assigned through visual inspection (Longtime/LT or Multiperiodic/MP) 
}
\end{table}

\clearpage

\bibliography{main}{}
\bibliographystyle{aasjournal}

\end{document}